%% file: main.tex
\documentclass[journal,12pt,onecolumn,draftclsnofoot,a4paper]{IEEEtran}
\usepackage[utf8]{inputenc}

\usepackage{multirow}
\usepackage{multicol}
\usepackage{url}

\input{Macros.tex}

\usepackage[scaled=0.3]{helvet}

\begin{document}

\title{A New Path to Code-based Signatures via Identification Schemes with Restricted Errors\thanks{The work of Edoardo Persichetti and (partially) Paolo Santini is supported by the National Science Foundation under Grant No. CNS-1906360.}}

\author{\IEEEauthorblockN{
Marco Baldi\IEEEauthorrefmark{1},
Massimo Battaglioni\IEEEauthorrefmark{1},
Franco Chiaraluce\IEEEauthorrefmark{1},\\
Anna-Lena Horlemann\IEEEauthorrefmark{2},
Edoardo Persichetti\IEEEauthorrefmark{3},\\
Paolo Santini\IEEEauthorrefmark{1} and
Violetta Weger\IEEEauthorrefmark{4}}\\
\IEEEauthorblockA{\IEEEauthorrefmark{1} Dipartimento di Ingegneria dell'Informazione, Università Politecnica delle Marche, Ancona, Italy}\\
\IEEEauthorblockA{\IEEEauthorrefmark{2} Faculty of Mathematics and Statistics, University of St. Gallen, Switzerland}\\
\IEEEauthorblockA{\IEEEauthorrefmark{3} Florida Atlantic University, Boca Raton, USA}\\
\IEEEauthorblockA{\IEEEauthorrefmark{4} Institute of Mathematics, University of Zurich, Zurich, Switzerland}
}

\maketitle

\begin{abstract}
In this paper we introduce a variant of the Syndrome Decoding Problem (SDP), that we call Restricted SDP (R-SDP), in which the entries of the searched vector are defined over a subset of the underlying finite field.  
We prove the NP-completeness of R-SDP, via a reduction from the classical SDP, and describe  algorithms which solve such new problem.
{We study the properties of random codes under this new decoding perspective, in the fashion of traditional coding theory results, and assess the complexity of solving a random R-SDP instance.}
As a concrete application, we describe how Zero-Knowledge Identification (ZK-ID) schemes based on SDP can be tweaked to rely on R-SDP, and show that this leads to compact public keys as well as significantly reduced communication costs.
Thus, these schemes offer an improved basis for the construction of code-based digital signature schemes derived from identification schemes through the well-know Fiat-Shamir transformation.
\end{abstract}

\section{Introduction} 
\label{sec:introduction}
\input{Introduction}

\section{Preliminaries}\label{sec:preliminaries}
\input{Preliminaries.tex}

\section{Zero Knowledge Identification Schemes based on syndrome decoding}\label{sec:previousZKID}
\input{PreviousZKIDschemes.tex}

\section{Decoding random codes with restricted errors}
\label{sec:restricted}
\input{Restricted}

\section{Solving R-SDP with maximum weight}
\label{sec:solving}
\input{Solving}

\section{Identification schemes based on R-SDP}
\label{sec:previous}
\input{Previous}

\section{Conclusion}\label{sec:concl}\input{Concl}

\newpage

\appendices

\section{The AGS scheme} 
\input{AppendixA}

\newpage
\section{Proof of proposition \ref{prop:wagner_1_complexity}}
\label{appendix_proof}
\input{AppendixB}

\newpage

\bibliographystyle{IEEEtran}

\bibliography{References}

\end{document}

%% file: Macros.tex
\usepackage{xcolor} 
\usepackage{amsfonts}
\usepackage{amsthm}
\usepackage{amsmath}
\usepackage{amssymb}
\usepackage{comment}
\usepackage{extarrows}
\usepackage{multirow}
\usepackage{enumerate}
\usepackage{array}
\usepackage{bigstrut}
\usepackage{exscale,relsize}
\usepackage{graphicx}
\usepackage{cite}
\usepackage[lined,ruled,linesnumbered,noend]{algorithm2e}
\SetAlCapSkip{1em}

\usepackage{tikz}
\usetikzlibrary{arrows,automata, positioning}
\usetikzlibrary{calc, chains,
                fit,
                positioning,
                shapes}
\usetikzlibrary{patterns}       

\usepackage{pgfplots}

\newcommand{\6}{\mathbf}

\newcommand{\rand}{\xleftarrow{\$}}

\newcommand{\pr}{\textsf{Pr}}

\newtheorem{prob}{Problem}
\newtheorem{prop}{Proposition} 
\newtheorem{proposition}[prop]{Proposition} 
\newtheorem{definition}[prop]{Definition} 
 
\newtheorem{theorem}[prop]{Theorem}

%% file: Introduction.tex
Public-key cryptography heavily relies on hard mathematical problems, that define the security target for the scheme by tying the system's private key to the public one, or a plaintext to the corresponding ciphertext.
Among the mathematical problems utilized in the context of public-key cryptography, one of the most studied is that of decoding random linear codes in the Hamming metric, which was proved to be NP-complete \cite{Berlekamp1978, barg1994some} and boasts a vast literature of algorithms aimed at finding a solution\cite{prange, dumer, leebrickell, leon, stern88, bjmm, mmt,ballcoll}. 
Roughly speaking, this problem asks to find a vector of low Hamming weight, i.e., containing only a few non-null entries, such that its product with a given parity-check matrix returns a target vector, called syndrome. For this reason, the problem is known as Syndrome Decoding Problem (SDP).

For decades, SDP has represented the foundation of the area called code-based cryptography. Indeed, it allows building efficient and secure cryptosystems, which are largely inspired by the seminal work of McEliece~\cite{McEliece1978}. However, the situation is not the same for signature schemes, and the vast majority of attempts to build a code-based signature scheme that is at the same time secure and efficient were unsuccessful. This is the case, for instance, of the scheme proposed by Courtois, Finiasz, and Sendrier~\cite{cfs}, denoted as CFS in the following, and its variants, which have tried to address the problem of decoding a random-like syndrome into a low-weight vector in several ways, but always yielding to either unpractical performance or even security breaches.
The recent work of~\cite{subsetsumSDP} highlights the fact that, over non-binary finite fields, the decoding problem remains hard also when the solution is required to have very high Hamming weight, as opposed to very low. Wave~\cite{debris2019wave} is a signature scheme built upon this fact, characterized by a public key size that grows as $\lambda^2$, where $\lambda$ is the security level in bits.
This provides an important improvement over CFS, although Wave still requires a public key of 3 megabytes for 128 bits of classical security, which is rather large.
All these classic code-based signature schemes rely on some hidden code structure, which is at the basis of the public key security and must be protected from attackers.

Constructing signatures via the Fiat-Shamir transform applied to a Zero-Knowledge Identification (ZK-ID) scheme is a promising alternative to this approach.
In fact, code-based identification schemes do not require any hidden structure, and hence are intrinsically resistant to structural attacks.
However, schemes such as Stern's~\cite{SternZK} feature non-trivial soundness errors and require many repetitions, leading to very large signature sizes, whereas attempts to translate the Schnorr-Lyubashevsky approach~\cite{lyubashevsky2012lattice}, like in~\cite{persichetti}, have been shown vulnerable to attacks based on statistical analysis~\cite{santiniattack, Deneuville2019}.  
Nevertheless, the scheme proposed by Cayrel, V{\'e}ron and El Yousfi Alaoui \cite{Cayrel2010}, denoted as CVE, has been shown suitable for obtaining very fast signature schemes with a reduction in the key size in the order of $25\%$ over Stern's \cite{ElYousfi2013}.\medskip

\paragraph*{\textbf{Our contribution}}

We introduce a variant of the SDP, in which the solution vector must take values over a restricted set of the finite field in which both the given code and the syndrome are defined.
For this reason, we denote the corresponding decoding problem as Restricted SDP (R-SDP), and prove its NP-completeness through a reduction from the classical SDP in the Hamming metric, over the same finite field. 
We then focus on a special case of R-SDP, in which the coordinates of the error vector are chosen from $\{0,\pm 1\}$, and use classical arguments from coding theory to derive conditions under which, for a random code, the solution of the problem is unique with overwhelming probability. 
{We assess the complexity of solving R-SDP, by taking into account recent works that focus on SDP over a ternary finite field \cite{Bricout}, and derive tight estimates for the resulting cost.}
As a culminating development of our work, we revisit the CVE scheme with a new formulation based on restricted errors.  
Our results show that, using such an approach, we can obtain a noticeable performance improvement by significantly reducing the communication cost and the public key size, while preserving the security level.

\medskip

\paragraph*{\textbf{Outline of the paper}} 
The paper is organized as follows. In Section \ref{sec:preliminaries} we introduce the notation that will be used throughout the paper and recall some coding theory notions. 
In Section \ref{sec:previousZKID} we recall the CVE scheme, 
which we use as the starting point of our variant.
In Section \ref{sec:restricted}, we introduce the concept of restricted error vectors and study their properties from a coding-theoretic point of view; we define the associated R-SDP, and prove its NP-completeness with a reduction from the canonical SDP. 
In Section \ref{sec:solving} we analyze different strategies to solve the R-SDP, in the specific case in which the weight of the searched vector is maximal; in particular, we take into account both brute force approaches and more clever strategies, which we adapt from recent results holding for the SDP over a ternary finite field \cite{Bricout}.
In Section \ref{sec:previous}, we adapt the CVE scheme to our new framework, and compare its performance to that of schemes in the existing literature. Finally, we draw some concluding remarks in Section \ref{sec:concl}.

%% file: Preliminaries.tex
In this section we introduce the notation that will be used throughout the paper, and provide some basic notions from coding theory.

\subsection{Notation}
We denote by $[a;b]$ the set of integers between $a$ and $b$ including $a$ and $b$. As usual, $\mathbb F_q$ denotes the finite field with $q$ elements, where $q$ is a prime power, while $\mathbb F_q^* = \mathbb F_q\setminus \{0\}$ denotes the  multiplicative group of $\mathbb F_q$.
We use bold upper case (resp. lower case) letters to denote matrices (resp. vectors).
For a matrix $\mathbf{A}$, we refer to its entry in the $i$-th row and $j$-th column as $a_{i,j}$, and for a vector $\mathbf{a}$ we denote its $i$-th entry by $a_i$.
The $k \times k$ identity matrix is denoted by $\mathbf{I}_k$. 
Let $\mathfrak S_n$ be the symmetric group on $n$ elements, where we  will represent elements $\sigma$ of $\mathfrak S_n$ as bijections from the integer set $[0;n-1]$ to itself, and the action on a length $n$ vector is represented as
$$\sigma(\6a) = \left(a_{\sigma(0)}, a_{\sigma(1)},\cdots,a_{\sigma(n-1)}\right).$$

We will use $\mathfrak M_n$ to denote the set of monomial transformations, i.e., all linear transformations that can be represented through the action of a permutation and non-zero scaling factors. In other words, for each $\tau\in \mathfrak M_n$, there exist $\sigma\in \mathfrak S_n$ and $\6v\in (\mathbb F_q^{*})^n$ such that
$$\tau(\6a) =  \left(v_{\sigma(0)}a_{\sigma(0)},v_{\sigma(1)}a_{\sigma(1)},\cdots,v_{\sigma(n-1)}a_{\sigma(n-1)}\right),\hspace{2mm}\forall \6a\in \mathbb F_q^n.$$

We use $\mathcal U(A)$ to denote the uniform distribution over a set $A$; for a random variable $a$, we write $a\sim \mathcal D$ if $a$ is distributed according to the distribution $\mathcal D$, and $a\xleftarrow{\$}A$ if $a$ is sampled according to the uniform distribution over $A$, i.e., 
$a\sim\mathcal U(A)$. 

The support of a vector $\mathbf{a}\in \mathbb{F}_q^n$  is defined as
$\mathrm{Supp}(\mathbf{a})= \{j\in [0 ; n-1]\hspace{2mm}|\hspace{2mm}a_j\neq 0\}$. For a set $J \subset [0; n-1]$ and a vector $\mathbf{a} \in \mathbb{F}_q^n$, we denote by $\mathbf{a}_J$ the vector consisting of the entries of $\mathbf{a}$ indexed by $J$. Analogously, for a set $J \subset [0;n-1]$ and a matrix $\mathbf{A} \in \mathbb{F}_q^{k \times n}$, we denote by $\mathbf{A}_J$ the matrix consisting of the columns of $\mathbf{A}$ indexed by $J$.

\subsection{Coding theory preliminaries}

Here we briefly recall some basic notions for codes in the Hamming metric.

\begin{definition}
An $[n,k]$ \emph{linear code} $\mathcal C$ over $\mathbb{F}_q$ is a linear subspace of $\mathbb{F}_q^n$ of dimension $k$. 
\end{definition}

Any linear code can be represented by a generator matrix, which has the code as image, or equivalently by a parity-check matrix, which has the code as kernel.

\begin{definition}
The \emph{Hamming weight} of $\mathbf{x} \in \mathbb{F}_q^n$  is equal to the size of its support, i.e.,
$$ \mathrm{wt}_{\mathrm H} (\mathbf{x}) := \mid \mathrm{Supp}(\mathbf{x})\mid = \mid \{ j \in [0; n-1] \mid x_j \neq 0 \} \mid. $$
The \emph{Hamming distance} between $\mathbf{x}$ and  $\mathbf{y} \in \mathbb{F}_q^n$   is defined as the Hamming weight of their difference, i.e.,
$$ d_\mathrm{H}\{\mathbf{x}, \mathbf{y} \} :=  \mathrm{wt}_\mathrm{H}(\mathbf{x}- \mathbf{y}) =\mid \{ i \in [0; n-1] \mid x_i \neq y_i \} \mid. $$
The \emph{minimum Hamming}   \emph{distance} of a code is the minimum of all pairwise non-zero Hamming  distances of the codewords.

\end{definition}

The sphere of vectors of $\mathbb F_q^n$ with Hamming weight $\omega$ is denoted by $\mathrm{S}^{\mathrm{H}}_{n,q,\omega}$ and the ball of radius $\omega$ is denoted by $\mathrm{B}^{\mathrm{H}}_{n,q,\omega}$.

In the following proposition we recall the Gilbert-Varshamov  bound in the Hamming metric.

\begin{proposition}[Theorem 13.73, \cite{berlekampbook}]\label{gv}
Let $q$ be a prime power, $n$ and $d_{\mathrm H}$ be positive integers. There exists a linear  code $\mathcal C$ over $\mathbb{F}_q$  of length $n$ and minimum Hamming distance $d_\mathrm{H}$, such that
\begin{equation*}
 \mid \mathcal{C} \mid \geq  \frac{q^n }{\sum_{j=0}^{d_\mathrm{H}-1}\binom{n}{j}(q-1)^j}.
 \end{equation*}
\end{proposition}

\begin{definition}
Let $q$ be a prime power and $0<k\leq n$ be positive integers.
For a code over $\mathbb F_q$ of length $n$ and dimension $k$, the \emph{Gilbert-Varshamov distance} is defined as follows
$$d_{\sf{GV}} := \max\left\{d_{\mathrm H}\hspace{1mm}\left|\hspace{1mm} \sum_{i=0}^{d_{\mathrm H}-1} \binom{n}{i} (q-1)^i < q^{n-k}\right.\right\}.$$
\end{definition}
It is well known \cite[Ch. 17, Problem (31)]{sloane} that random codes in the Hamming metric asymptotically attain the Gilbert-Varshamov distance with overwhelming probability.

Finally, we formally state the Syndrome Decoding Problem.

\begin{prob}\label{prob:sdp}\textbf{Syndrome Decoding Problem (SDP)}
Given $\6H\in\mathbb F_q^{(n-k)\times n}$, $\6s\in\mathbb F_q^{n-k}$ and $t\in\mathbb N$, decide whether there exists $\6e\in\mathbb F_q^n$, with $\mathrm{wt}_{\mathrm H} (\mathbf{e})\leq t$, such that $\6e\6H^\top = \6s$. 
\end{prob}

As mentioned in the introduction, the SDP is the foundation of code-based cryptography. In particular, when the Hamming weight $t$ is below the error correction capability following from the Gilbert-Varshamov bound, the SDP has at most one solution with overwhelming probability, when $q$ and $n$ are large.

%% file: PreviousZKIDschemes.tex
In this section we focus on the CVE scheme \cite{Cayrel2010}. This scheme will be described in detail in Section \ref{sec:CVE}, since it is the scheme to which we apply our new technique (see Section \ref{sec:previous}).
For the sake of completeness, in Appendix A we also provide a description of a scheme by Aguilar, Gaborit and Schrek \cite{Aguilar2011}, which we will denote by AGS and whose performance will be compared to that of the scheme we propose.

Let us briefly recall the operating principles of such schemes, highlighting their main features; then, in the following sections, we show how switching to restricted errors leads to a strong boost in their performance.

\subsection{General principles}

Let $\sf{R}$ be a relation which is satisfied only by specific pairs of objects, such that checking whether a pair of elements satisfies the relation is efficient (i.e., it can be done in polynomial time).
An identification scheme constructed upon $\sf{R}$ can be defined as a two-stage procedure, as follows:
\begin{enumerate}
    \item[-] in the first stage, the \emph{prover} randomly generates a pair $(\sf{sk}, \sf{pk})$ satisfying $\sf{R}$;
    \item[-] in the second stage, the prover exchanges messages with the \emph{verifier}, which is only equipped with $\sf{pk}$, with the goal of demonstrating knowledge of $\sf{sk}$. At the end of the protocol, the verifier decides whether to accept the prover or not. 
\end{enumerate}
Usually, the key pair is such that $\sf{pk}$ represents an instance of a hard problem, with $\sf{sk}$ being a valid solution.
Thus, the difficulty of finding, on input $\sf{pk}$, a value $\sf{sk}^*$ that satisfies the relation, without knowledge of $\sf{sk}$, is at the core of the scheme, since authentication is obtained through the proof of knowledge about the secret key.
\medskip

An identification scheme is called \emph{zero-knowledge} if no information about the secret key is revealed during the identification process.  
Other required properties for an identification scheme are \emph{completeness} and \emph{soundness}, the former meaning that a honest prover always gets accepted and the latter requiring that an impersonator has only a small probability of getting accepted. 
For a rigorous and complete description of these properties, we refer the interested reader to \cite{galbraith2017identification}. 
\medskip

\paragraph*{\textbf{Reducing the communication cost}}
A crucial quantity to analyze in an identification scheme is the \emph{communication cost}, i.e., the cost of a full interaction between the two parties, which is measured as the number of bits that are exchanged.
Many identification schemes are constructed from Sigma protocols, i.e., three-pass proofs of knowledge for a certain relation. In this case, the scheme presents a 
\emph{soundness error}, meaning that an adversary impersonating a prover (without access to the secret key) can ``cheat'' by pre-selecting a candidate response that works only for a subset of the challenge space, and hope that the chosen challenge is part of that subset. This implies that an impersonator is able to get accepted with a certain non-zero probability (e.g., 2/3 or 1/2), which depends on the scheme. It follows that, in order to achieve an acceptable level of authentication, the protocol is repeated several times, and a prover is accepted only if every instance was answered successfully. If the cheating probability is $\eta$, executing $N$ rounds of the protocol leads to an overall authentication level of $\eta^{N}$, and $N$ is chosen so that the desired value (e.g., $2^{-128}$), is achieved.

\begin{figure}[ht!]
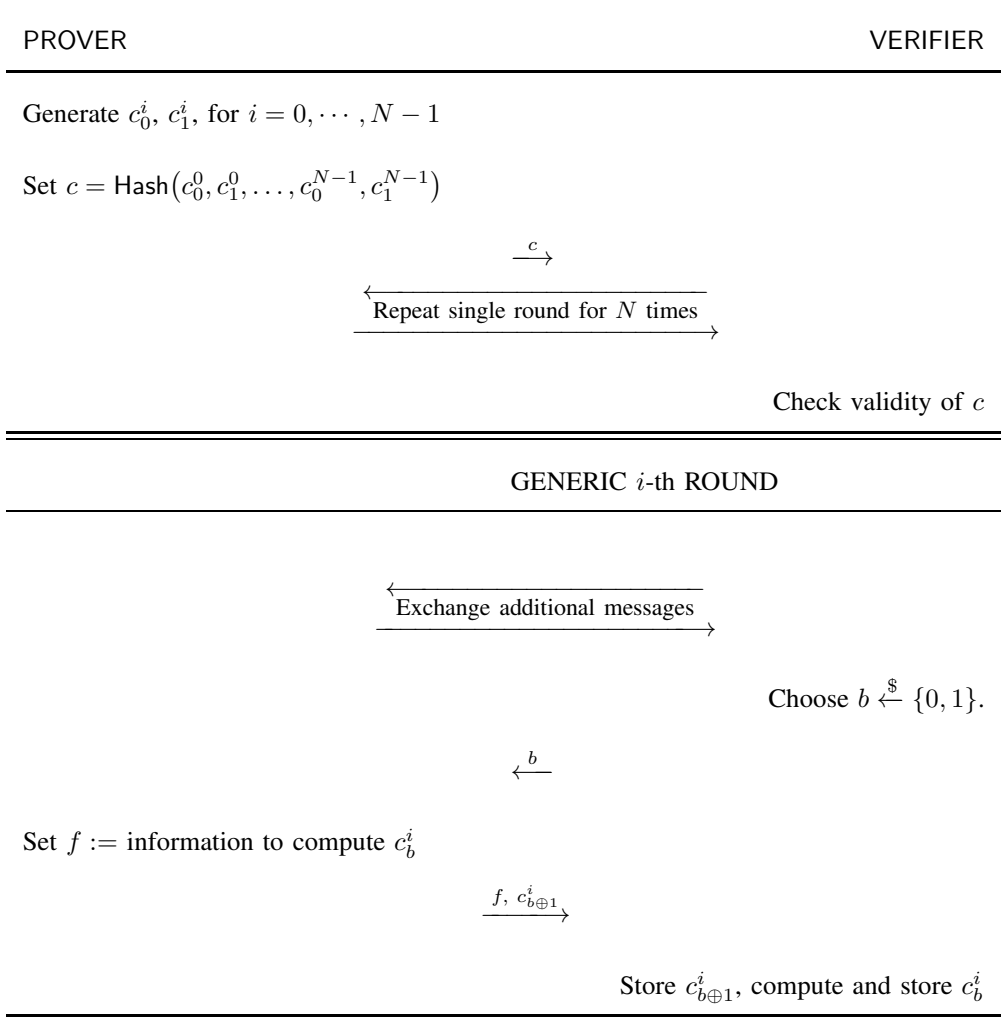
\small
\centering
\begin{tabular}{p{6cm}p{1cm}p{4.8cm}}
\hline
\textsf{PROVER} & & \multicolumn{1}{r}{\textsf{VERIFIER}}\\
\hline
Generate $c_0^i$, $c_1^i$, for $i=0,\cdots,N-1$  &  & \\
Set $c = \textsf{Hash}\big( c_0^0, c_1^0,\ldots, c_0^{N-1}, c_1^{N-1} \big)$ &   & \\
 & $\xlongrightarrow{c}$  &\\
& \makebox[\dimexpr(\width-10mm)][l]{\hspace{-21mm}$\xlongrightarrow{\xlongleftarrow[\text{\footnotesize Repeat single round for $N$ times}]{}}$} &\\
&&  \multicolumn{1}{r}{Check validity of $c$}\\
\hline
\hline
 & \mbox{GENERIC $i$-th ROUND} & \\
 \hline
 &&\\
& \makebox[\dimexpr(\width-10mm)][l]{\hspace{-18mm}$\xlongrightarrow{\xlongleftarrow[\text{\footnotesize Exchange additional messages}]{}}$} &\\
 &&\multicolumn{1}{r}{Choose $b\xleftarrow{\$}\{0 , 1\}$.}\\
&$\xlongleftarrow{b}$&\\
Set $f:=$ information to compute $c_b^i$ &&\\ 
& \makebox[\dimexpr(\width-10mm)][l]{\hspace{-4mm}$\xlongrightarrow{f, \hspace{1mm} c_{b\oplus 1}^i}$} & \\
&&\multicolumn{1}{r}{Store $c^i_{b\oplus 1}$, compute and store $c^i_b$}\\
\hline
\end{tabular}
\caption{Description of the compression technique for $N$ rounds.}
\label{fig:cayrelcompression}
\end{figure}

In order to illustrate the process, we will analyze the case of the CVE scheme, that is the main focus of our work. In this scheme, each round is based on the following paradigm:
\begin{enumerate}
    \item the prover prepares two commitments $c_0$, $c_1$, which are obtained on the base of some randomness;
    \item the two commitments are sent to the verifier; after this exchange, some additional messages may be exchanged between the two parties;
    \item the verifier randomly picks $b\in\{0 , 1\}$, and sends it to the prover;
    \item the prover provides information that only allows to verify $c_b$, but not $c_{b\oplus 1}$;
    \item the verifier checks the validity of $c_b$.
\end{enumerate}
When the protocol is repeated for multiple rounds, it is possible to reduce the overall communication cost by exploiting the compression technique proposed in \cite{Aguilar2011}. 
For the sake of completeness, we summarize this procedure in Fig.~\ref{fig:cayrelcompression}. 
Before the $0$-th round, the prover generates the commitments for all the $N$ rounds, and then sends a unique hash value $c = \textsf{Hash}\big( c_0^0, c_1^0,\ldots, c_0^{N-1}, c_1^{N-1} \big)$ to the verifier. 
In the $i$-th round, after receiving the challenge $b$, the prover sets its response $f$ such that the verifier can compute $c^i_b$, and additionally includes $c_{b\oplus 1}^i$.  
At the end of each round, the verifier uses $f$ to compute $c^i_b$, and stores it together with $c_{b\oplus 1}^i$. 
After the final round only, the verifier is thus able to check validity of the initial commitment $c$, by computing the hash of all the stored $c^i_0$, $c^i_1$. 
This way, one hash is sent at the beginning of the protocol, and only one hash (instead of two) is transmitted in each round: this way, the number of exchanged hash values reduces from $2N$ to $N + 1$. 
For the sake of clarity, this compression technique will not be included in the description of the forthcoming schemes, but we remark that 
this can be applied, with slight modifications, to all the schemes we analyze in the rest of the paper.

\renewcommand{\arraystretch}{1.5}

\subsection{The CVE scheme}\label{sec:CVE}

\begin{figure}[ht!]
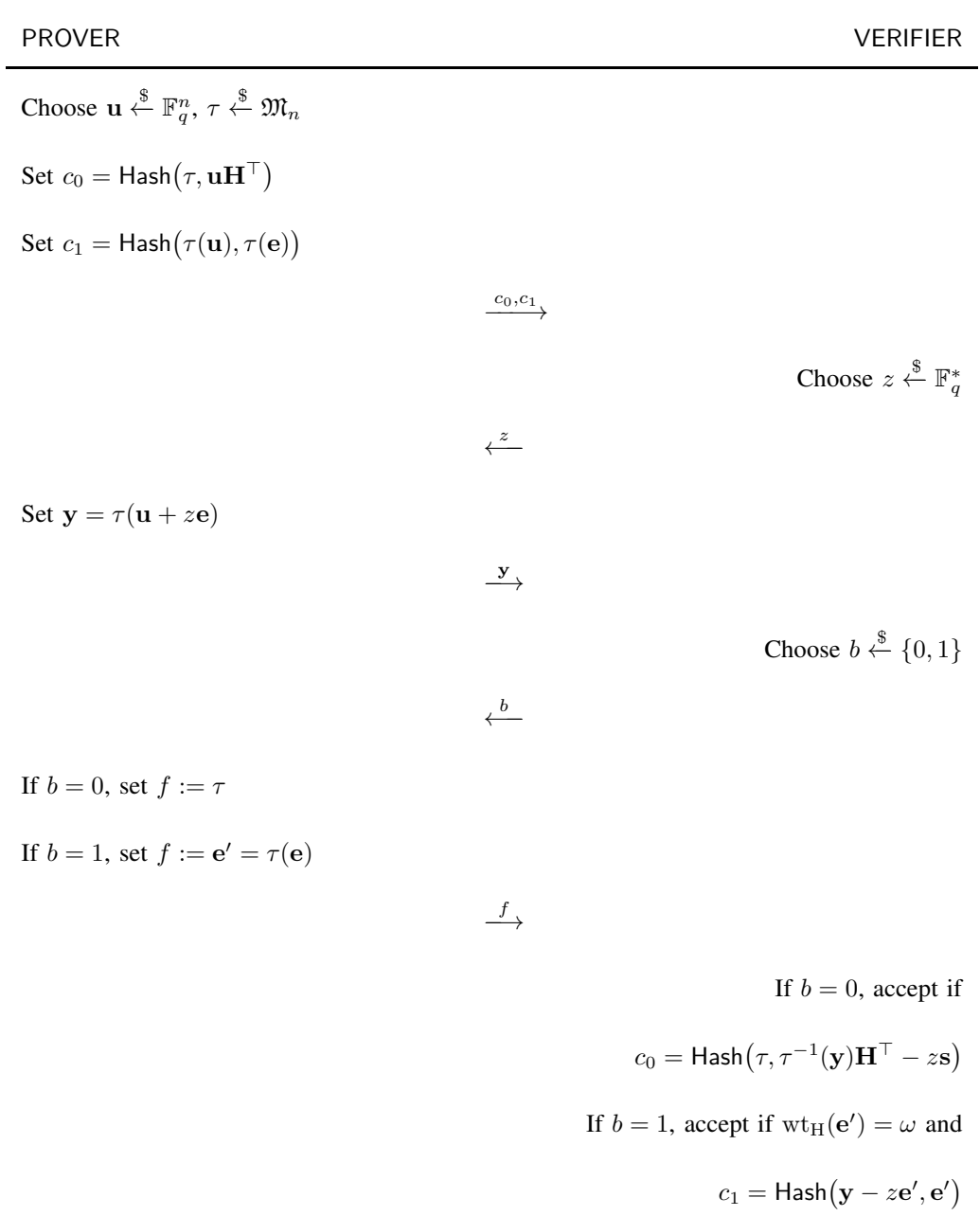
\small
\centering
\begin{tabular}{p{6cm}p{1cm}p{4.8cm}}
\multicolumn{3}{l}{\textsf{Public Data}\quad Parameters $q,n,k,\omega\in\9N$, parity-check matrix $\6H\in\mathbb F_q^{(n-k) \times n}$}\\
\multicolumn{3}{l}{\textsf{Private Key}\quad $\6e\in \mathrm{S}^{\mathrm H}_{n,q,\omega}$}\\
\multicolumn{3}{l}{\textsf{Public Key}\quad\ $\6s = \6e\6H^\top\in\mathbb F_q^{n-k}$} \\[5pt]
\hline
\textsf{PROVER} & & \multicolumn{1}{r}{\textsf{VERIFIER}}\\
\hline
Choose $\6u\xleftarrow{\$}\mathbb F_q^n$, $\tau\xleftarrow{\$} \mathfrak M_n$  &  & \\
Set $c_0 = \textsf{Hash}\big(\tau,\6u\6H^\top\big)$  &  &\\
Set $c_1 = \textsf{Hash}\big(\tau(\6u), \tau(\6e)\big)$  &  &\\
 & $\xlongrightarrow{c_0,c_1}$  &\\
 &&\multicolumn{1}{r}{Choose $z\xleftarrow{\$}\mathbb F_q^*$}\\
&$\xlongleftarrow{z}$&\\
Set $\6y =  \tau(\6u+z\6e)$ &&\\
&$\xlongrightarrow{\6y}$&\\
& & \multicolumn{1}{r}{Choose $b\rand \{0,1\}$}\\
& $\xlongleftarrow{b}$ &\\
If $b=0$, set $f:=\tau$ & & \\
If $b=1$, set $f:=\6e' = \tau(\6e)$ & & \\
&$\xlongrightarrow{f}$&\\
&&\multicolumn{1}{r}{If $b=0$, accept if}\\
&&\multicolumn{1}{r}{$c_0 = \textsf{Hash}\big(\tau,\tau^{-1}(\6y)\6H^\top-z \6s\big)$}\\
&&\multicolumn{1}{r}{If $b=1$, accept if $\mathrm{wt}_{\mathrm H}(\6e') = \omega$ and }\\
&&\multicolumn{1}{r}{$ c_1 = \textsf{Hash}\big(\6y - z\6e',\6e'\big)$}\\
\hline
\end{tabular}
\caption{The CVE scheme.}
\label{fig:cayrelcompre}
\end{figure}

The CVE scheme \cite{Cayrel2010} is an improvement of Stern's \cite{SternZK} and V\'eron's \cite{Veron97} identification schemes, both based on the hardness of decoding a random binary code \cite{Berlekamp1978}. 
The former scheme relies on non-binary codes over a large finite field. With this choice, the cheating probability for a single round is reduced from $2/3$ of Stern's 3-pass scheme to $\frac{q-1}{2q}$ by using a 5-pass scheme based on codes over $\mathbb{F}_q$.
For the sake of completeness, the CVE scheme is summarized in Fig.~\ref{fig:cayrelcompre}.
\renewcommand{\arraystretch}{1.5}
We now briefly recall how the communication cost of this scheme is derived \cite[Section 4.2]{Cayrel2010}.
We first note that, in order to represent a length-$n$ vector of weight $\omega$ over $\mathbb F_q$, we can either use the full vector, or just consider its support, together with the ordered non-zero entries.
The first option requires $n\left\lceil \log_2(q)\right\rceil$ bits, while for the second one we need $\omega\big(\left\lceil\log_2(n)\right\rceil+\left\lceil\log_2(q-1)\right\rceil\big)$ bits.
By considering the most convenient choice for each set of parameters $n$, $\omega$ and $q$, representing a length-$n$ vector of weight $\omega$ over $\mathbb F_q$ requires $\psi(n,q,\omega) = \min\{n\left\lceil \log_2(q)\right\rceil,\omega\big(\left\lceil\log_2(n)\right\rceil+\left\lceil\log_2(q-1)\right\rceil\big)\}$ bits. 
Furthermore, objects that have been randomly generated (such as the monomial transformations) can be compactly represented by the sole seed that is used as input of the pseudorandom generator.
Taking all of this reasoning and the compression technique into account, and denoting with $l_{\textsf{Hash}}$ and $l_{\textsf{Seed}}$ the length of hash values and seeds, respectively, for $N$ rounds of the protocol we get the following average communication cost: 
$$l_{\textsf{Hash}}+N \bigg(\left\lceil\log_2(q-1)\right\rceil+n\left\lceil\log_2(q)\right\rceil+1+l_{\textsf{Hash}}+\frac{\psi(n,q,\omega)+l_{\textsf{Seed}}}{2}\bigg).$$
For the maximal communication cost, we take the maximum size of the response, and thus we obtain
$$l_{\textsf{Hash}}+N \bigg(\left\lceil\log_2(q-1)\right\rceil+n\left\lceil\log_2(q)\right\rceil+1+l_{\textsf{Hash}}+\max\{\psi(n,q,\omega)\hspace{1mm},\hspace{1mm}l_{\textsf{Seed}}\}\bigg).$$

In order to derive secure parameters for the CVE scheme, one can proceed as follows.
For a given set of parameters $n$, $k$ and $q$, the Gilbert-Varshamov bound is used to estimate the minimum distance $d_{\mathrm H}$ of a random code described by an $(n-k) \times n$ parity-check matrix $\6H$ over $\mathbb{F}_q$.
The weight of the private key can then be set as $\omega = \left\lfloor d_{\mathrm{H}}/2\right\rfloor$, since this guarantees that there is no other vector of weight smaller than or equal to $\omega$ with syndrome equal to the public key.
In order to reach a security level of $\lambda$ bits, $\omega$ must be sufficiently large, such that using the best known attack algorithms requires a number of operations not lower than $2^{\lambda}$.
The authors of \cite{Cayrel2010} have used the analysis due to Peters \cite{PetersFq} to estimate the Information Set Decoding (ISD) complexity, and have proposed two parameters sets:
\begin{enumerate}
    \item[-] $q=256$, $n=128$, $k=64$, $\omega=49$, for 87-bits security;
    \item[-] $q=256$, $n=208$, $k=104$, $\omega=78$, for 128-bits security. 
\end{enumerate}

%% file: Restricted.tex
In this section we introduce a variant of the decoding problem, in which the error vector is constrained to take values in a subset of the finite field; for this reason, we speak of \emph{restricted errors}. 
In particular, we show that the associated decoding problem is NP-complete and adapt the Gilbert-Varshamov bound to this case. We also study the complexity of solving the decoding problem with restricted errors, through adaptions of modern algorithms for the Hamming case.

\subsection{R-SDP and NP-completeness}\label{subsec:NP}

In this section we introduce a new variant of the decoding problem, by choosing a set of vectors over $\mathbb F_q$, with $q$ being an odd prime power, whose Hamming weight is below some threshold value; as an additional restriction, the vectors in the set take values in a specific subset of the finite field.
For the  elements of the multiplicative group $\mathbb F_q^*$ associated to $\mathbb F_q$, 
we adopt a representation $\mathbb F_q^* = \left\{x_1 = 1,x_2,\cdots,x_{q-1}\right\}$, such that $x_i+x_{q-i} = 0$, for $i\in [0 ; \frac{q-1}{2}]$. 
We will call this representation for the finite field elements ``symmetric"; it is evident that, for the same finite field, many symmetric representations may exist. 
Note that, when $q$ is a prime, the canonical symmetric representation is $\mathbb F_q^* = \left\{1,2,\ldots,q-1\right\}$.
\medskip

More precisely, for a positive integer $a\leq\frac{q-1}{2}$, that we call the \emph{restriction parameter}, we define the restricted Hamming ball of radius $t$ and parameter $a$ as
$$E_{n,q,t}^{(a)} := \left\{\6e\in\mathbb F_q^{n} \mid \mathrm{wt}_{\mathrm{H}}(\6e) \leq t,\hspace{2mm}\6e \in \{0,\pm x_1,\cdots,\pm x_a\}^n\right\}.$$ 
Hence,  we are focusing on a subset of $\mathbb{F}_q$ of size $2a+1$ where, for each element, the set also contains its additive inverse.
We formally define the syndrome decoding problem for a restricted Hamming ball over an arbitrary finite field as follows.
\begin{prob}\textbf{Restricted Syndrome Decoding Problem (R-SDP)}
Let $q = p^m$, with $p\neq 2$ being a prime and $m\in\mathbb N$, and denote by $\mathbb F_q$ the corresponding finite field with $q$ elements, described through a symmetric representation. 
On input $\6H\in\mathbb F_q^{(n-k)\times n}$, $\6s\in\mathbb F_q^{n-k}$ and $t\in\mathbb N$, decide whether there exists an $\6e\in E^{(a)}_{n,q,t}$, such that $\6e\6H^\top = \6s$. 
\end{prob}
The above problem is obtained by applying an additional restriction to the classical SDP over a finite field $\mathbb F_q$. It may thus seem intuitive that, like SDP, R-SDP belongs to the hierarchical class of NP-complete problems.
A formal proof of this property is provided in the following theorem.
Note that, in our proof, the restriction parameter $a$ is not treated as an input to R-SDP, since  we can prove the NP-completeness for any fixed value of $a$.

\begin{theorem}
R-SDP is NP-complete.
\end{theorem}

\begin{IEEEproof}
We provide a reduction from the classical SDP in the Hamming metric, defined over $\mathbb F_q$, as formulated in Problem~\ref{prob:sdp}. This was proved to be NP-complete in \cite{barg1994some}.

Clearly, the finite field representation does not interfere with the definition of the problem; thus, we directly consider a symmetric representation for the field.
From now on, for the sake of simplicity, we will denote $\mathbb F_q^{(a)} = \{0,\pm x_1, \ldots, \pm x_a\}$.

We denote by $\{\6H,\6s,t\}$ an arbitrary instance of SDP, and map it into an R-SDP instance, that we denote by $\{\6H',\6s,t\}$. 
If $q=3$, then we can set $\6H' = \6H$; otherwise, we construct $\6H'$ according to the following procedure.
We first select a set $U\subseteq \mathbb F_q^*$, such that every element of $\mathbb F_q^*$ can be obtained as the product of one element from $\mathbb F_q^{(a)}$ and one from $U$, that is
$$\forall b\in\mathbb F_q^*\hspace{2mm} \exists u\in U,\hspace{1mm}x\in\mathbb F_q^{(a)}\hspace{2mm}\text{such that}\hspace{2mm}-x u = b\hspace{2mm}\text{or}\hspace{2mm}x u = b.$$
It is easily seen that, for all possible sets $\mathbb F_q^{(a)}$, a choice for $U$ always exists and for its cardinality, that we denote by $v$, it is straightforward to show that  $\lfloor \frac{q-1}{2a} \rfloor \leq v\leq q-1$. 
It may also happen that, for the same element of $\mathbb F_q^*$, more than one pair of factors from $U$ and $\mathbb F_q^{(a)}$ exists but, for our purposes, this is not an issue.
Let $\6u=(u_0,\ldots,u_{v-1})$ be a vector formed by the elements of $U$: we finally obtain $\6H'$ as 
\begin{center}
    $\6H'=\Big[\begin{array}{cccc}
\6u\otimes \6h_0 & \6u\otimes \6h_1& \cdots  & \6u\otimes \6h_{n-1}\\[5pt]
\end{array}\Big]$,
\end{center}
where $\otimes$ denotes the Kronecker product 
and $\6h_i$ is the $i$-th column of $\6H$. 
This way, $\6H'$ has $n-k$ rows and $n' = nv$ columns, given by
$$\6h'_{iv+j} = u_j  \6h_i,\hspace{2mm}i\in [0 ; n-1],\hspace{2mm}j\in [0 ; v-1].$$

We first show that, to each $\6e'\in\big(\mathbb F_q^{(a)}\big)^{n'}$ such that $\6e'\6H'^\top = \6s$, we can associate a vector $\6e\in\mathbb F_q^n$ such that $\6e\6H^\top = \6s$.
In fact, consider that, $\forall i \in [0; n-1]$, 
$$\sum_{j=0}^{v-1}e'_{iv+j} \6h'_{iv+j} = \left(\sum_{j=0}^{v-1}e'_{iv+j}  u_j\right) \6h_{i} = \beta_i \6h_i,$$
where $\beta_i\in\mathbb F_q$.
On the one hand, it is easily seen that, if $e'_{iv+j} = 0$ for all $j\in[0 ; v-1]$, then $\beta_i = 0$ as well. For the other cases, it might happen that $\beta_i\neq 0$.
The number of coefficients $\beta_i$ that are non-zero is surely smaller than or equal to $\mathrm{wt}_{\mathrm{H}}(\6e')$ and, if $\6s\neq \60_r$, then at least one of the $\beta_i$ must be non-zero.
Let $\6e\in\mathbb F_q^n$, such that $e_i = \beta_i, \forall i\in[0;n-1]$: it is then immediately seen that $\6e\6H^\top = \6s$, and $\mathrm{wt}_\mathrm{H}(\6e)\leq \mathrm{wt}_\mathrm{H}(\6e')\leq t$.
If $\6e'\in E^{(a)}_{n',q,t}$, then $\6e$ has weight smaller than or equal to $t$: existence  of an $\6e'$ satisfying the constructed R-SDP instance, then, implies existence of an $\6e$ satisfying the initial SDP instance.

Finally, we show that $\6e'\in E^{(a)}_{n',q, t}$ exists if and only if a desired $\6e$ exists, i.e., if the initial SDP instance is a ``yes" instance.
In fact, as a consequence of the requirements on $U$, for each $\beta_i$ we can always write 
$$\beta_i = \lambda_i u_{\ell_i} x_{y_i},$$ 
for proper indices $\ell_i\in[0 ; v-1]$, $y_i\in[1 ; a]$ and $\lambda_i\in\{\pm 1\}$.
Then, we can always build a vector $\6e'\in E^{(a)}_{n',q, t}$ having at most one non-zero entry among the ones in positions $\{iv,\cdots,iv+v-1\}$, whose elements are defined as
$$e'_{iv + j} = \begin{cases} 
\lambda_i x_{y_i} & \text{if $j=\ell_i$},\\
0 & \text{elsewhere}.\end{cases}$$
This vector has the same weight as $\6e$, and is such that $\6e\6H^\top = \6e'\6H'^\top$.
Given the bijection between $\6e$ and $\6e'$, it becomes clear that solving the given R-SDP instance means solving the initial SDP instance.
It follows that R-SDP is NP-complete.

\end{IEEEproof}\medskip

We have a two-fold motivation to study R-SDP, arising from our interest in identification schemes based on syndrome decoding.
Firstly, in these schemes, the proof of knowledge is provided by publishing a masked version of the secret key, which is a low-weight vector over $\mathbb F_q$.
As already stressed in the previous sections,  a crucial quantity to study the performance of an identification scheme is the communication cost per round. 
Vectors with restricted entries can clearly be represented with a lower number of bits, compared to vectors over the full field $\mathbb F_q$. Thus, we expect the performance of these schemes to benefit from the use of restricted vectors.
For this reason, from now on we limit our attention to the case of $a = 1$, which leads to the lowest communication cost, since we impose the maximal non-trivial restriction to the subset $B$. Secondly, these schemes are parameterized by choosing the weight of the secret error vector as large as possible, but in such a way that only few solutions exist.
We then study the cost of the best known algorithms to solve the SDP adapted to this scenario. 
For a random code, the minimum Hamming distance is estimated via the Gilbert-Varshamov bound, which only depends on the code length, dimension and on the finite field size.
We generalize all these concepts to the case of restricted vectors, and derive conditions to guarantee that, for a given random code, only a few 
solution to the  R-SDP exist. 
Our results show that, for random codes with the same parameters, we can achieve higher security levels by relying on R-SDP instead of SDP.
As we show more extensively in Section \ref{sec:previous}, our results lead to strong improvements in the performances of already existing identification schemes.

\subsection{Restricted minimum distance and properties of random codes}\label{subsec:prop}

In this section we study the properties of random linear codes regarding the weight of codewords with restricted entries.
As mentioned before, we will focus on the case $a=1$ and, to ease the notation, we will simply use $E_{n,q,t}:=E_{n,q,t}^{(1)}$ to denote the ball of vectors with entries in $\{0,\pm 1\}$ and with Hamming weight not larger than $t$.
Since the sum of two vectors in $E_{n,q,t}^{(1)}$ results in a vector in $E_{n,q,t}^{(2)}$, we also need to study the latter ball. 
Our goal is to derive conditions upon which, for a given code over $\mathbb F_q$, with $q \geq 5$ being an odd prime power, the R-SDP has at most a unique solution with overwhelming probability. 

\begin{definition}
Let $q \geq 5$ be an odd prime power, and denote by $\mathbb F_q$ the corresponding finite field with $q$ elements, described with a symmetric representation.
For $\6a\in\mathbb F_q^n$, we define the \emph{ restricted weight} of $\6a$  as 
$$\widetilde{\mathrm{wt}}(\6a) := \begin{cases}
\#_1(\6a) + 2\cdot\#_2(\6a) & \text{if $\6a\in\{0,\pm 1, \pm 2\}^n$,} \\ \infty & \text{otherwise,}
\end{cases}$$
where $\#_1(\6a)$ is the number of entries of $\6a$ equal to $\pm 1$ and $\#_2(\6a)$ is the number of entries equal to $\pm 2$. 
\end{definition}

Note that, since $q$ is a prime, the  restricted weight of a vector over $\{0,\pm 1, \pm 2\}$ corresponds to its Lee weight~\cite{lee1958some}. 
\begin{definition}Let $q \geq 5$ be an odd prime power, and denote by $\mathbb F_q$ the corresponding finite field with $q$ elements, described with a symmetric representation.
Let $\mathcal C$ be a linear code with length $n$ over $\mathbb F_q$. We define its \emph{restricted minimum  distance} as
$$\widetilde d := \min\left\{\widetilde{\mathrm{wt}} (\6c)\mid \6c\in\mathcal C\cap\{0,\pm 1, \pm 2\}^n\setminus \{\60_n\}\right\}. 
$$
If $\mathcal C\cap\{0,\pm 1, \pm 2\}^n= \{\60_n\}$, then we set $\widetilde d =\infty$.
\end{definition}

Note that the corresponding restricted distance is not a proper metric, but a pre-metric, and that the restricted minimum distance of $\mathcal C$ is equal to the minimum Lee distance of $\mathcal{C} \cap \{0, \pm 1, \pm2 \}^n$.
Its importance is stated in the following theorem. 
\begin{theorem}\label{the:unique_syn}
Let $q \geq 5$ be an odd prime power, and denote by $\mathbb F_q$ the corresponding finite field with $q$ elements, described with a symmetric representation.
Let $\mathcal C$ be a code over $\mathbb F_q$, with length $n$ and  restricted minimum distance $\widetilde{d}$.
For any parity-check matrix $\6H$ for $\mathcal C$, and for all $t< \widetilde{d}/2$, there cannot exist two distinct vectors $\6e,\6e'\in E_{n,q,t}$ such that $\6e\6H^\top = \6e'\6H^\top$.
\end{theorem}
\begin{IEEEproof}
If $\6e\6H^\top = \6e'\6H^\top$, then necessarily $\tilde{\6e} = \6e-\6e'\in\mathcal C$. Furthermore, given that $\6e,\6e'\in\{0,\pm 1\}^n$ and $\6e \ne \6e'$, we have $\tilde{\6e}\in\{0,\pm 1, \pm 2\}^n$ and 
$\widetilde{\mathrm{wt}}(\tilde{\6e})\leq 2t$. This contradicts the fact that the restricted minimum distance of $\mathcal C$ is $\widetilde d>2t$. 
\end{IEEEproof}
Note that, if a code has no codewords (apart the zero one) living in $\{0,\pm 1, \pm 2\}^n$, then its  restricted minimum distance is infinite and all vectors over $\{0,\pm 1\}$ correspond to distinct syndromes.

Analogously to the Hamming metric, we define a ball of radius $t$ and center $\6a\in\mathbb F_q^n$ as the set of all vectors whose difference with $\6a$ has  restricted weight smaller than or equal to $t$, that is
\begin{equation}\label{rest_ball}
\widetilde B(\6a,t,q,n) := \left\{\6x\in\mathbb F_q^n\mid \widetilde{\mathrm{wt}}(\6x-\6a)\leq t\right\}.
\end{equation}
The volume of each such  ball does not depend on its center but only on its radius, hence we get the following.
\begin{proposition}
Let $q \geq 5$ be an odd prime power, and $n$ and $t\leq 2n$ be positive integers. Then the size of a  restricted ball in $\mathbb F_q^n$ 
of radius $t$,  as defined in \eqref{rest_ball}, is given by 
$$\widetilde{V}(n,t) := \sum_{i=0}^{t}\sum_{j=\max\{0,i-n\}}^{\left\lfloor i/2 \right\rfloor}\binom{n}{j}\binom{n-j}{i-2j}2^{i-j}.$$
\end{proposition}
Note that this is the same as the size of the Lee ball of radius $t$ in $\mathbb{F}_5^n$, which can be found in \cite[Proposition 8, Corollary 9]{LeeNP}.

In the following theorem we derive a bound which, in the same fashion of the Gilbert-Varshamov bound, states the minimal maximum  dimension for a code achieving a given  restricted minimum distance.
\begin{theorem}
For a given finite restricted minimum distance $\widetilde{d}$ and length $n$,  there exists a code in $\mathbb F_q^n$ of dimension $\tilde k$, where 
\begin{align*}\tilde k \nonumber & \geq n - 1 -  \log_q{\left(\widetilde{V}(n,\widetilde{d}-1)\right)}\\\nonumber
& = n - 1 -  \log_q{\left(\sum_{i=0}^{\widetilde{d}-1}\sum_{j=\max\{0,i-n\}}^{\left\lfloor i/2 \right\rfloor}\binom{n}{j}\binom{n-j}{i-2j}2^{i-j}\right)}.
\end{align*}
\end{theorem}
\begin{IEEEproof}
Let $\mathcal C\subseteq \mathbb F_q^n$ with  restricted minimum distance $\widetilde{d}$ and maximum dimension $\tilde k$. In the following we show that for every vector $\6x\in\mathbb F_q^n$, there must be at least an $\alpha_{\6x}\in\mathbb F_q^*$ and a codeword $\6c$ in $\mathcal C$ such that $\widetilde{\mathrm{wt}}(\alpha_{\6x} \6x-\6c)\leq \widetilde{d}-1$ or, analogously, $\alpha_{\6x}\6x \in \widetilde B(\6c,\widetilde{d}-1,q,n)$.
In other words, for each $\6x\in\mathbb F_q^n$, at least one of its scalar multiples is contained in the ball of radius $\widetilde{d} - 1$ with center in a codeword of $\mathcal C$.
Note that, if $\6x\in\mathcal C$, then this is trivially true since it is enough to consider a sphere with center in $\6x$; to show that this holds for all vectors in the space, we consider $\6x\not\in\mathcal C$.
If there is no pair $(\alpha_{\6x},\6c)$ satisfying the above requirement, then we can consider a new code $\mathcal C'$, defined as
$$\mathcal C' = \left\{\beta \6x+\6c\mid
\beta\in\mathbb F_q,\hspace{2mm}\6c\in\mathcal C\right\}.$$ 
Such a code will be linear and will have dimension $\tilde k+1$. Furthermore, its restricted minimum distance will not be lower than that of $\mathcal C$, because by hypothesis all ``new" codewords $\beta \6x+\6c$, with $\beta\neq 0$, are outside of balls of radius $\widetilde{d}-1$ centered in codewords of $\mathcal C$.
However, the existence of $\mathcal C'$ contradicts the fact that, by hypothesis, $\mathcal C$ has maximum dimension: thus, it must be
$$\forall \6x\in\mathbb F_q^n\hspace{1mm}\exists \alpha_{\6x}\in\mathbb F_q^*,\hspace{1mm}\6c\in\mathcal C\hspace{2mm}\text{s.t.}\hspace{2mm}\alpha_{\6x} \6x\in \widetilde{B}(\6c, \widetilde d - 1, q, n).$$ 
Let $A$ be the set of all valid vectors $\alpha_{\6x} \6x$: for a given $\6x$, we include all the scalar multiples such that the above condition is verified.
It is easily demonstrated that
$$q^{n-1}\leq \mid A  \mid \leq q^n.$$
In fact, $A\subseteq \mathbb F_q^n$ proves the upper bound, while the lower bound is obtained by assuming that for each $\6x$, only one scalar multiple satisfies the condition. 
Given that $A=\bigcup_{\6c\in\mathcal C}\widetilde{B}(\6c,\widetilde{d} -1, q, n)$, we consider the following chain of inequalities
\begin{align*}
q^{n-1} \leq |A| = \left|\bigcup_{\6c\in\mathcal C}\widetilde{B}(\6c,\widetilde{d}-1, q , n)\right|\leq\sum_{\6c\in\mathcal C}\left|\widetilde{B}(\6c,\widetilde{d}-1, q , n)\right| = q^{\tilde k}  \widetilde{V}(n,\widetilde{d} -1).
\end{align*}
Simple further computations yield the claimed inequality. 
\end{IEEEproof}\medskip

Starting from the previous theorem, in the same fashion of the commonly called Gilbert-Varshamov distance for the Hamming metric, we define the \emph{restricted Gilbert-Varshamov minimum distance}.
\begin{definition}
For a code with length $n$ and dimension $k$ over $\mathbb F_q$, with $q\geq 5$ being an odd prime power, we define the \emph{restricted Gilbert-Varshamov distance} as follows
$$\widetilde{d}_{\sf{GV}} := \begin{cases}\infty & \text{if $\widetilde{V}(n,2n)<q^{n-k-1}$,}\\
\max\left\{\widetilde{d}>0\hspace{1mm}\left|\hspace{1mm} \widetilde{V}(n,\widetilde{d})\leq q^{n-k-1}\right.\right\}&\text{otherwise.}\end{cases}$$
\end{definition}

Analogously to the classic formulations, the restricted Gilbert-Varshamov distance tells us the maximum distance that a code of fixed length and rate can achieve. 
Note that $\widetilde{V}(n,2n) = 5^n$, and hence for large values of $q$ and/or $n$ (if $q>5$), the restricted Gilbert-Varshamov distance is equal to $\infty$, as long as $k<n(1-\log_q(5)-\frac{1}{n})$. 
Thus, it makes sense to study the probability with which a random code achieves this restricted minimum distance; in the next theorem we show that this probability is bounded from below by a quantity that, as the code length increases, asymptotically tends to 1.
\begin{theorem}\label{the:final_GV_probability}
Let $q>5$ be an odd prime power and $k \leq n\left(1-\log_q(5)-\epsilon\right)$, for $0 < \epsilon < 1-\log_q(5)$.
Let $\6G\xleftarrow{\$}\mathbb F_q^{k\times n}$ with rank $k$.
Then the code generated by $\6G$ has restricted  minimum distance $\widetilde{d} = \infty$ with probability at least
$1 - q^{-\epsilon n}$.

\end{theorem}
\begin{IEEEproof}
First, the requirements on $k$ and $\epsilon$ assure that $0<k<n$.
The code generated by $\6G$ will have restricted minimum distance $\widetilde{d} = \infty$ if all its codewords (apart from the zero one) do not live in $\{0,\pm 1, \pm 2\}^n$.
We first focus on a single codeword: since the entries of $\6G$ are random over $\mathbb F_q$, each linear combination of the rows of $\6G$ is a random length-$n$ vector with entries over $\mathbb F_q$, as well.
Thus, the probability that this codeword has a finite restricted weight (i.e., it takes values in $\{0,\pm 1 ,\pm 2\}$) is
$$\frac{\left|\{0,\pm 1, \pm 2\}^n\setminus \60_n\right|}{\left|\mathbb F_q^n\setminus \60_n\right|} = \frac{5^n-1}{q^n-1} < \frac{5^n}{q^n} = q^{-n\left(1-\log_q(5)\right)}.$$
Since there are $q^k$ codewords, from a union bound argument we get that the probability that the code contains at least a codeword of restricted weight smaller than or equal to $2n$ is at most
$$q^k q^{-n(1-\log_q(5))} \leq q^{n\left(1-\log_q(5)-\epsilon\right)} q^{-n(1-\log_q(5))} = q^{-\epsilon n }.$$
Then, $q^{-\epsilon n}$ is an upper bound on the probability that the code generated by $\6G$ contains at least a codeword with finite weight; taking its complement, we obtain a lower bound on the desired probability.
\end{IEEEproof}

Note that, in Theorem \ref{the:final_GV_probability}, we need to assume $q>5$; in fact, any non-zero code in $\mathbb F_5^n$ must contain non-zero codewords from $\{0,\pm 1,\pm 2\}^n=\mathbb F_5^n$ and hence the restricted minimum distance will always be finite.

%% file: Solving.tex
In this section we describe how R-SDP can be solved, in the particular case in which the  restricted weight of the searched vector is maximal (i.e., equal to the code length). 
We focus on this particular case since it is the one we consider for our modification of the CVE scheme, which we introduce in the next section.
We chose this restriction since, as we show in the next section, it is the one yielding the best performance in our target application.
Note that one can also rely on ISD algorithms to solve R-SDP; however, when the weight is maximal, then ISD algorithms collapse to the brute force approach described in Section \ref{bruteforce}.
Indeed, since the searched vector contains no zero elements, the whole procedure of an ISD is reduced to searching for all possible vectors in the chosen information set.
For solving the R-SDP with arbitrary restricted weight, we note that one can use any ternary ISD, e.g., Peter's generalization \cite{PetersFq} of Stern's algorithm \cite{stern88}. The only difference to this algorithm is that the operations are performed not over $\mathbb{F}_3$ but over $\mathbb{F}_q$, thus we assume that addition has a cost of  $\left\lceil \log_2(q) \right\rceil$ and multiplication has a cost of  $\left\lceil \log_2(q) \right\rceil^2$. 

Let us consider a random code with length $n$ and dimension $k$, described by $\6H\in\mathbb F_q^{(n-k)\times n}$, 
and $\6s\in\mathbb F_q^{n-k}$, where
$\6s\rand\left\{\6H\6e^\top\mid \6e\in\{\pm 1\}^n\right\}$.
In other words, we assume that the given R-SDP instance has at least one solution.
\medskip

We start by determining the number of solutions that exist on average.
\begin{prop}\label{prop:num_solutions}
Let $\6H\in\mathbb F_q^{(n-k)\times n}$ be the parity-check matrix of a random code with rate $R = k/n$.
Let $\6s\rand\left\{\6e\6H^\top\mid \6e\in\{\pm 1\}^n\right\}$, and let us denote the set of length-$n$ vectors with entries over $\{\pm 1\}$  and syndrome equal to $\6s$ as $\mathcal M$.
Then, unless $n$ is very small, the average cardinality of $\mathcal M$ is
\begin{equation}
M \approx  1+2^{n\big(1-(1-R)\log_2(q)\big)}.
\label{eq:cardM}
\end{equation}
\end{prop}
\begin{IEEEproof}
The thesis results from simple combinatorial arguments.
First, we know that the existence of at least one solution is guaranteed. 
Let $\6e\in\{\pm 1\}^n$ such that $\6s = \6e\6H^\top$, and $V = \{\pm 1\}^n\setminus \6e$.
Since we are considering a random code, we have that $\6H$ is a random matrix, hence sums of its columns will yield random vectors over $\mathbb F_q^{n-k}$. 
Thus, for each $\6e'\in V$, we can assume that $\6e'\6H^\top$ is random over $\mathbb F_q$; therefore, the probability that the product equals $\6s$ is given by $q^{-(n-k)}$.
Hence, on average, the cardinality of $\mathcal M$ is
$$M = 1+\frac{|V|}{q^{n-k}} = 1 + \frac{2^n-1}{q^{n-k}}\approx 1 +  \frac{2^n}{q^{n-k}} = 1 + 2^{n\big(1-(1-R)\log_2(q)\big)};$$
notice that the approximation in this expression holds unless $n$ is really small.
\end{IEEEproof}

Note that this result implies that for growing $q$ and/or $n$ there will be most likely only one solution, which concurs with the result in Theorem \ref{the:final_GV_probability}.

We now proceed by describing how the R-SDP can be solved in the considered setting.
We first consider a basic brute force approach, and then take into account the analysis in \cite{Bricout} to derive a more efficient approach based on merging.

\subsection{Brute force approach}\label{bruteforce}

Given $\6H$, we can choose a permutation $\pi\in\mathfrak S_n$ and a full rank matrix $\6S\in\mathbb F_q^{(n-k)\times(n-k)}$ such that $\6S\pi(\6H) = [\6H' \quad \6I_{n-k}]$, where $\pi(\6H)$ denotes the matrix obtained by permuting the columns of $\6H$ according to $\pi$, and $\6H'\in\mathbb F_q^{(n-k)\times k}$.
Notice that, if $\6e\6H^\top = \6s$, then $\pi(\6e)$ is going to be a solution to the R-SDP instance defined by the parity-check matrix $\6S\pi(\6H)$ and syndrome $\6s' = \6S\6s$.
Let us write $\6x = (\6x_0,\6x_1)$, with $\6x_0\in\{\pm 1\}^k$ and $\6x_1\in\{\pm 1\}^{n-k}$. 
Then we want the following condition to be verified
$$\6x_0\6H'^\top+\6x_1 = \6s',$$
from which $\6x_1 = \6s'-\6x_0\6H'^\top$.
Thus, to solve the R-SDP, it is enough to go through vectors $\6x_0\in\{\pm 1\}^{k}$ and, for each candidate, compute the corresponding $\6x_1$ and test whether it has entries over $\{\pm 1\}$. 
As soon as such a vector is found, we have a solution to the initial R-SDP, that is, $\pi^{-1}\big([\6x_0, \6x_1]\big)$.
\medskip

We can estimate the complexity of the above approach as follows.
\begin{prop}\label{prop:brute_complexity}
Let $(\6H,\6s)$, with $\6H\in\mathbb F_q^{(n-k)\times n}$ drawn at random with rank $n-k$, and $\6s\rand\left\{\6e\6H^\top\mid\6e\in\{\pm 1\}^n\right\}$.
With the brute-force approach, one can solve the R-SDP represented by $(\6H,\6s)$ with the following average cost 

$$\frac{(n-k)^2(n+1)\left\lceil\log_2(q)\right\rceil^2}{\prod_{j = 1}^{n-k} \left(1-q^{-j} \right)} + \frac{2^k}{1+M}\cdot\frac{q}{q-2} k \lceil \log_2(q) \rceil,$$
where $M = 1+2^{n\big(1-(1-R)\log_2(q)\big)}$ is the average number of solutions.  
\end{prop}

\begin{IEEEproof}
We neglect the cost of the column permutation.
It is immediately seen that $ \prod_{j = 1}^{n-k} \left(1-q^{-j} \right)$ corresponds to the probability that a random  matrix of size $n-k$ and entries over $\mathbb F_q$ is non singular; given that $\6H$ is random, $ \prod_{j = 1}^{n-k} \left(1-q^{-j} \right)$ also corresponds to the probability that $\pi(\6H)$ can be put into systematic form.
To do this, we use Gaussian elimination, with an overestimating 
cost given by $(n-k)^2(n+1)\left\lceil\log_2(q)\right\rceil^2$.
Once we have transformed both $\6H$ and $\6s$, we start by testing vectors $\6x_0\in\{\pm 1\}^k$. 
For each vector, we compute the corresponding $\6x_1$, with a cost of $k\left\lceil\log_2(q)\right\rceil$.  
For this we consider early abort, i.e., we test simultaneously if the entries of $\6x_1$ are in $\{ \pm 1\}$. A random element in $\mathbb{F}_q$ has probability $\frac{q-2}{q}$ to not be in $\{\pm 1\}$, thus on average we can abort after computing $\frac{q}{q-2}$ many entries. 
The expected number of existing solutions is $M = 1 + 2^{n\big(1-(1-R)\log_2(q)\big)}$ (see Proposition \ref{prop:num_solutions}). 
Thus, we have $M$ vectors $\6x_0$ leading to a solution: on average, before we find one solution, we test $2^k/(1+M)$ candidates for $\6x_0$.
\end{IEEEproof}

\subsection{Using the PGE+SS framework}

We notice that the R-SDP, for $q=3$, is essentially identical to the so-called ternary syndrome decoding problem, i.e., the problem of decoding a random linear code defined over $\mathbb F_3$.
For such a problem, the state-of-art solver is described in \cite{Bricout}, and follows the so-called Partial Gaussian Elimination plus Subset Sum (PGE+SS) framework, briefly recalled in the following.
We directly adapt such a framework to our case, and parameterize the whole description of the algorithm on an integer $\ell\in[1 ; n-k]$, whose meaning will be clarified next.
The PGE+SS approach is characterized by the following four consecutive steps:
\begin{enumerate}
    \item[1) ] \emph{Permutation}: pick a random $\pi\in\mathfrak S_n$ and apply $\pi(\6H)$ by permuting the columns of $\6H$ according to $\pi$.
    \item[2) ]\emph{PGE}: divide $\pi(\6H)$ into four blocks, that is
    $$\pi(\6H) = \begin{bmatrix}
    \6H'_0 \in\mathbb F_q^{(n-k-\ell)\times (n-k-\ell)}& \6H'_1\in\mathbb F_q^{(n-k-\ell)\times (k+\ell)}\\
    \6H'_2\in\mathbb F_q^{\ell\times (n-k-\ell)} & \6H'_3\in\mathbb F_q^{\ell\times (k+\ell)}
    \end{bmatrix}.$$
    If $\6H'_0$ is singular, return to step 1, otherwise reduce $\pi(\6H)$, i.e., find $\6S\in\mathbb F_q^{(n-k)\times (n-k)}$, such that
    $$\6S\pi(\6H) =  \begin{bmatrix}
    \6I_{n-k-\ell}& \6H'\in\mathbb F_q^{(n-k-\ell)\times (k+\ell)}\\
    \60_{\ell\times (n-k-\ell)} & \6H''\in\mathbb F_q^{\ell\times (k+\ell)}
    \end{bmatrix}.$$
    Write $\6s\6S = [\6s',\6s'']$, 
    with $\6s'\in\mathbb F_q^{n-k-\ell}$ and $\6s''\in\mathbb F_q^{\ell}$.
\item[3) ]\emph{Small R-SDP}: produce a set $\mathcal E\subseteq \{\pm 1\}^{k+\ell}$ containing (some) solutions to the R-SDP instance represented by $\6H''$ and $\6s''$, i.e., such that
$$\6s'' = \6H''\6e''^\top,\hspace{2mm}\forall \6e''\in\mathcal E.$$
\item[4) ]\emph{Test}: for each $\6e''\in\mathcal E$, test whether $\6e' = \6s'-\6e''\6H'^\top$ has entries over $\{\pm 1\}$. If such a vector $\6e'$ is found, return $\pi^{-1}\big([\6e',\6e'']\big)$; otherwise, restart from step 1.
\end{enumerate}

We now briefly explain the rationale behind the procedure.
First, for each $\6e\in\{\pm 1\}^n$ such that $\6s = \6e\6H^\top$, it also holds that 
$\6s\6S = \pi(\6e)\big(\6S\pi(\6H)\big)^\top$.
Let us write $\pi(\6e) = [\6e' , \6e'']$; then
\begin{equation}
\label{eq:small}
\begin{cases}\6e'+\6e''\6H'^\top = \6s',\\
\6e''\6H''^\top = \6s''.
\end{cases}    
\end{equation}
Notice that the pair $(\6H'',\6s'')$ corresponds to another R-SDP instance, since the unknown $\6e''$ has length $k+\ell$ and  restricted weight $k+\ell$.
This instance is characterized by a smaller size, since $\6H''$ defines a code with length $k+\ell\leq n$ and dimension $k$; since $\6H$ is random, $\6H''$ is random as well.
Let $\mathcal M$ denote the set of solutions to the initial R-SDP instance: for each vector in $\mathcal M$, an associated solution to the small instance exists.
Namely, let $\mathcal M'$ be the set formed by the last $k+\ell$ entries of the vectors obtained by permuting those in $\mathcal M$, according to $\pi$: each vector in $\mathcal M'$ is a solution for the small instance.
Thus, every time $\mathcal E$ and $\mathcal M'$ are not disjoint, we have that each vector in their intersection can be used to build a solution for the initial R-SDP instance.
The rationale of this procedure is in the fact that, depending on $\ell$, building solutions to the small instance may be significantly easier. However, we remark that the value of $\ell$ must be carefully optimized: indeed, when it is too low, the small instance may be pretty easy to solve but, as a side-effect, it may have too many solutions and testing them may become rather time consuming.
\medskip

In the following section we describe how the small R-SDP instance can be solved relying on Wagner's algorithm   \cite{wagner}.
In order to assess the complexity of this approach, we study the algorithm under the assumption that it succeeds when it finds one of the vectors in $\mathcal M'$.
In other words, if we find a vector $\6e''\in\{\pm 1\}^{k+\ell}$ such that $\6e''\6H''^\top = \6s''$ but $\6e''\not\in\mathcal M'$, then we do not consider this as a valid solution.
We assume that the cardinality of $\mathcal M'$ is the same as that of $\mathcal M$, and we estimate it through \eqref{eq:cardM}.
Notice that this is a conservative assumption because, in principle, it may happen that the cardinality of $\mathcal M'$ is smaller than that of $\mathcal M$.
Indeed, when we permute the vectors in $\mathcal M$ and consider only their last $k+\ell$ entries, it may happen that two (or more) of the vectors obtained are identical.

\subsection{Solving the small instance with Wagner's algorithm}

As in \cite{Bricout}, we rely on Wagner's algorithm to solve the small instance, which can be rewritten in the form of a subset sum problem, where the elements of the given set correspond to the columns of $\6H''$ and can be seen as elements of the finite field with $q^\ell$ elements. 
Let us consider a general description of Wagner's algorithm and analyze its complexity.
We anticipate the conclusion that, differently from the case of the ternary SDP studied in \cite{Bricout}, in our case the most convenient approach is to rely on Wagner's algorithm structured on just one level.
\medskip

The structure of Wagner's algorithm depends on a positive integer $a$, which defines the number of levels upon which the algorithm is divided.  
The set $[0;k+\ell-1]$ is partitioned into $2^a$ sets $J_i$ of approximately the same size $(k+\ell)/{2^a}$. 
Without loss of generality, we can assume that the entries of each set $J_i$ are consecutive, i.e., $J_i = \left\{\left\lfloor i\cdot \frac{k+\ell}{2^a}\right\rfloor, \ldots,  \left\lfloor (i+1)\cdot \frac{k+\ell}{2^a}\right\rfloor-1\right\}$, for $i \in [0 ; 2^{a}-1]$. 
For the sake of simplicity, we assume that $2^a$ divides $k+\ell$: thus, all sets $J_i$ have the same size $\frac{k+\ell}{2^a}$.
Let $\6H''_{j}$ denote the $\ell\times \frac{k+\ell}{2^a}$ matrix formed by the columns of $\6H''$ that are indexed by $J_i$.
\medskip

We start on level 0 by choosing random subsets $\mathcal R_0, \dots, \mathcal R_{2^a-1}$ from $\{\pm 1\}^{\frac{k+\ell}{2^a}}$, each with size $2^{v}$, where $v$ is arbitrary, and building the initial lists 
\begin{align}
\label{eq:initial}
\mathcal L^{(0)}_j = \left\{\left.\big(\6z = \6p\6H''^\top_j\hspace{1mm},\hspace{1mm}\6p\big)\right|\6p\in \mathcal{R}_j \right\},\hspace{2mm}\text{for $j\in[0;2^a-2]$}, \\
\mathcal L^{(0)}_{2^{a}-1} = \left\{\left.\big(\6z = \6p\6H''^\top_{2^a-1} -\6s''\hspace{1mm},\hspace{1mm}\6p\big)\right|\6p\in \mathcal{R}_{2^a-1} \right\}.
\end{align}

The algorithm then proceeds, in each level, by pairwise merging the input lists, as explained in detail below, 
and using the resulting lists as the input for the subsequent level; this procedure halts when, in the $a$-th level, one remains with only one list.
In each level, the number of input lists equals $2^{a-i+1}$ 
and the number of the produced lists is $2^{a-i}$.
In order to merge the lists, the algorithm uses  $a-1$ positive integers $0<u_1<u_2< \ldots < u_{a-1} < \ell$; 
furthermore, we set $u_0 = 0$ and $u_a = \ell$.
In order to merge the lists in the $i$-th level, for $i \in [1;a]$ we use the following procedure:
    from two lists $\mathcal L^{(i)}_{2j}$ and $\mathcal L^{(i)}_{2j+1}$, with $j<2^{a-i}-1$,  
    we produce the list 
    \begin{align*} \mathcal{L}_j^{(i+1)} & =  \mathcal L^{(i)}_{2j} \sqcap_{u_i} \mathcal L^{(i)}_{2j+1}  \\
    & = \{ (\6z_{2j}+\6z_{2j+1}, [\6p_{2j}, \6p_{2j+1}]) \mid (\6z_b, \6p_b) \in \mathcal{L}_{b}^{(i)}, \6z_{2j} + \6z_{2j+1} = \60 \ \text{in the last} \ u_i \ \text{entries} \}.
    \end{align*}
 
When the level $a$ is reached, only one list remains, containing vectors $\6p\in\{\pm 1\}^{k+\ell}$ such that $\6p\6H''^\top = \6s''$.

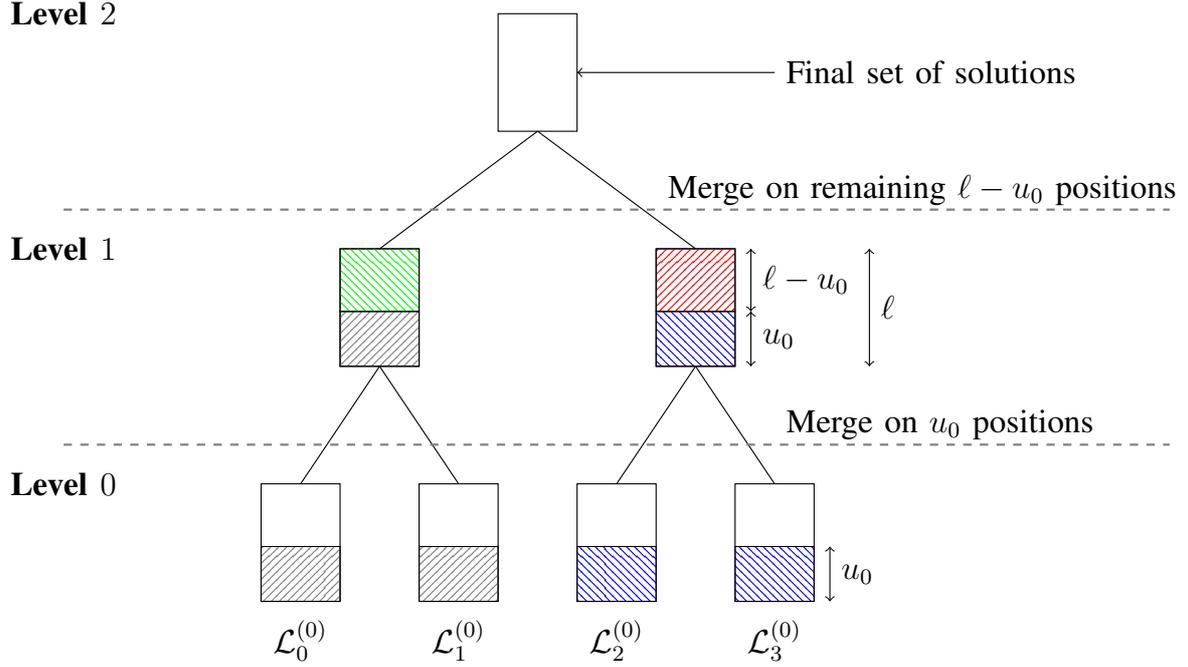
\begin{figure}
    \centering
    \input{wagner}
    \caption{Wagner's algorithm structured on two levels.}
    \label{fig:wagner}
\end{figure}

The whole procedure is detailed in Algorithm \ref{wagner}, 
while a graphical description of the algorithm, for the case of $a=2$, is shown in Figure \ref{fig:wagner}.
\medskip
\begin{algorithm}[ht!]\label{wagner}
\KwIn{$\mathbf{H''}_0,\cdots,\6H''_{2^a-1} \in \mathbb{F}_{q}^{\ell \times \frac{k+\ell}{2^a}}$, $\mathbf{s''}\in\mathbb{F}_{q}^{\ell}$, $v\in \mathbb N$ such that $v\leq \frac{k+\ell}{2^a}$, $a-1$ positive integers $u_1 < \cdots < u_{a-1} < \ell$.} 
\KwOut{A list $\mathcal L_0^{(a)} = \left\{\big( \6p\6H''^\top\hspace{1mm},\hspace{1mm}\6p\big)\right\}$ such that $\6p\in\{\pm 1\}^{k+\ell}$ and $\6p\6H''^\top = \6s''$} 
Set $u_0 = -1$, $u_a = \ell$.
\\Choose random subsets $\mathcal{R}_0,\cdots, \mathcal{R}_{2^a-1} \subseteq \{\pm 1\}^{(k+\ell)/2}$, each of size $2^v$.
\\Build the lists $\mathcal L_j^{(0)} = \left\{(\6z = \6p\6H''^\top_j, \6p)\mid \6p\in\mathcal{R}_j\right\}$ for $j \in [0; 2^a-2]$.
\\Build the list $\mathcal L^{(0)}_{2^{a}-1} = \left\{\left.\big(\6z = \6p\6H''^\top_{2^a-1} -\6s''\hspace{1mm},\hspace{1mm}\6p\big)\right|\6p\in \mathcal{R}_{2^a-1} \right\}$.
\\\For{$i = 1$    \emph{\textbf{to}} $a$}{
\For{$j = 0$ \emph{\textbf{to}} $2^{a-i}-2 $}{ 
$\mathcal L^{(i+1)}_j = \mathcal L^{(i)}_{2j} \sqcap_{u_i} \mathcal L^{(i)}_{2j+1} $} }
\Return $\mathcal L^{(a)}_0$
\caption{Wagner's algorithm structured on $a$ levels}
\end{algorithm}

Let us explain the rationale of the algorithm. For $\6x$ such that $\6x\6H''^\top = \6s''$, we indicate with $\6x_i$ the vector formed by the entries of $\6x$ that are indexed by $J_i$.
Notice that
\begin{equation}
\label{eq:wagner_decomposition}
\sum_{j = 0}^{2^a-1}\6x_j{\6H_j}''^\top = \6s''.
\end{equation}
Wagner's algorithm searches for solutions of the R-SDP instance $(\6H'', \6s'')$ by  exploiting the representation in \eqref{eq:wagner_decomposition}: one starts with lists of candidates for each sub-vector $\6x_j$ and, in each level, filters them to obtain, in the end, a set of solutions.
In the final step, i.e., on level $a$, we get a final list of solutions of the smaller R-SDP, since we look for vectors adding up to $\6s''$.
Since the lists are only chosen of size $2^v$, the algorithm is probabilistic.
We remind that we are interested in finding specific solutions to the small R-SDP instance, namely, we want Wagner's algorithm to find one of the vectors in $\mathcal M'$; moreover, we assume $|\mathcal M'| = M = 1+2^{n\big(1-(1-R)\log_2(q)\big)}$.
In the next proposition we derive the probability that Wagner's algorithm ends with a success.
\begin{proposition}\label{prop:pr_wagner}
We consider Wagner's algorithm based on $a$ levels, with parameters $0<u_1<\cdots<u_{a-1}<\ell$ and with initial lists of size $L_0 = 2^v$, with $v\leq\frac{k+\ell}{2^a}$. 
We assume that the cardinality of $\mathcal M'$ is $M =1+ 2^{n\big(1-(1-R)\log_2(q)\big)}$, and that each vector in $\mathcal M'$ is random over $\{\pm 1\}$.
Then, the probability of finding a valid solution is
$$1-\left(1-2^{v2^a-k-\ell-\delta\log_2(q)}\right)^M,$$ with
$\delta = \sum_{i = 1}^{a-1}u_i 2^{a-1-i}$.

\end{proposition}
\begin{IEEEproof}
Let $\6x\in\mathcal M'$ be one of the solutions.
First, Wagner's algorithm searches a solution over $\mathcal R_0\times \cdots \times\mathcal R_{2^a-1}\subseteq \{\pm 1\}^{k+\ell}$.
It is clear that whenever $\6x \not\in\mathcal R_0\times\cdots\times \mathcal R_{2^a-1}$ then $\6x$ will never be included in the final list.
Thus, the probability that $\6x$ actually is among the considered vectors is
$$\left(\frac{L_0}{2^{\frac{k+\ell}{2^a}}}\right)^{2^a} = 2^{v2^a-k-\ell}.$$
Notice that, even when $\6x$ is among the explored vectors, it may be filtered as a consequence of lists merging.
Indeed, let us divide $\6x$ into $2^a$ chunks, each formed by $\frac{k+\ell}{2^a}$ consecutive entries, which we denote as $\6x_j$, for $j \in [0;2^{a}-1]$.
Let us consider the merge in the $i$th level: $\6x$ will not be filtered in this level if and only if
\begin{enumerate}
\item[1. ] for $j \in[ 0;2^{a-i}-2]$, $\6x_{2j}\6H''^\top_{2j}+\6x_{2j+1}\6H''^\top_{2j+1}$ contains only zeros in the last $u_i-u_{i-1}$ positions;
\item[2. ] $\6x_{2^{a-i+1}-2}\6H''^\top_{2^{a-i+1}-2}+\6x_{2^{a-i+1}-1}\6H''^\top_{2^{a-i+1}-1} -\6s''$ contains only zeros in the last $u_i-u_{i-1}$ positions.
\end{enumerate}
Note that conditions 1 and 2 are actually not independent: indeed, it is easily seen that if condition 1 is met, then condition 2 is met as well. 
Given that both $\6H''$ and $\6x$ are random, in each merge, chunks $\6x_{2j}$ and $\6x_{2j+1}$ will not be filtered out with probability $q^{-u_i+u_{i-1}}$.
Given that, for $j\in [0 ; 2^{a-i}-2]$, we perform $2^{a-i}-1$ merges, condition 1 is verified with probability
$$\left(q^{-u_i+u_{i-1}}\right)^{2^{a-i}-1} = 2^{-(2^{a-i}-1)(u_i-u_{i-1})\log_2(q)}.$$
Hence, the above equation corresponds to the probability that $\6x$ is not filtered in the $i$th level.

Thus, the probability of $\6x$ surviving till the last level is
$$\prod_{i = 1}^{a-1}\left(q^{-(u_{i}-u_{i-1})}\right)^{2^{a-i}-1} = 2^{-\log_2(q)\sum_{i = 1}^{a-1}\left(2^{a-i}-1\right)(u_i-u_{i-1})},$$
where $u_0 = 0$.
With simple further computations, the above probability can be expressed as $2^{-\delta\log_2(q)}$, where $\delta$ is as in the claim.
In the end, the probability that a particular $\6x$ is i) explored through the initial lists, and ii) not filtered, is
$$2^{v2^a-k-\ell-\delta\log_2(q)}.$$
Since we assume that $\mathcal M'$ contains $M$ vectors, which are independent and random over $\{\pm 1\}$, we easily derive the probability of finding at least one of them as the complementary of the probability that we are not able to find any of these vectors.
\end{IEEEproof}

Roughly speaking, the complexity of Wagner's algorithm can be estimated with the maximum size of the produced lists.
Indeed, in order to merge two lists of some size, one can first join and sort the two lists and then proceed by finding pairs of vectors which sum up to zero in some prescribed coordinates. 
Neglecting polynomial or logarithmic factors, the complexity corresponds to the list size.
Let us now derive an expression for the list size growth in Wagner's algorithm.
In the initial level, we prepare lists of size $L_0 = 2^v$.
In the first level, i.e., for $i = 1$, the average size of the lists is given by $L_1 = L_0^2 / q^{u_1} = 2^{2v-u_1\log_2(q)}$. 
In the $i$-th level, for $i\geq 1$, the average size of the lists $\mathcal{L}_j^{(i)}$ is $L_{i} = \frac{L_{i-1}^2}{q^{u_{i}-u_{i-1}}}$.
Thus, for $i\geq 0$ we have

\begin{equation}
L_{i} = 2^{v2^{i}-\gamma(i)\log_2(q)},\hspace{2mm}\text{with\hspace{2mm}}
\gamma(i) = 
\begin{cases}
0&\text{if $i = 0$,}\\
u_{i}+\sum_{m = 1}^{i-1}2^{i-1-m}u_{m}&\text{otherwise.}
\end{cases}    
\end{equation}

Taking the maximum of these sizes, and dividing it by the probability that Wagner's algorithm finds a solution, we estimate its asymptotic complexity.
As mentioned above, list merging can be performed with a cost that (neglecting polynomial and logarithmic factors) corresponds to the list size (see Appendix B for more details on how this cost can be estimated in the finite regime).
Hence, a rough estimate of Wagner's algorithm complexity is as in the next proposition.
\begin{prop}
We consider Wagner's algorithm based on a number of levels equal to $a$, with parameters $0<u_1<\cdots<u_{a-1}<\ell$ and with initial lists of size $L_0 = 2^v$, with $v\leq\frac{k+\ell}{2^a}$. 
We assume that the cardinality of $\mathcal M'$ is $M =1+ 2^{n\big(1-(1-R)\log_2(q)\big)}$, and that $\mathcal M'$ is uniformly distributed in $\{\pm 1\}^{k+\ell}$.
Then, neglecting polynomial and logarithmic factors, we estimate the cost
of Wagner's algorithm as
$$\frac{\max_{i\in [1 ; a]}\left\{2^{v2^i-\gamma(i)\log_2(q)}\right\}}{1-\left(1-2^{v2^a-k-\ell-\delta\log_2(q)}\right)^M},$$
where $\delta = \sum_{i = 1}^{a-1}u_i 2^{a-1-i}$. 
When $M2^{v2^a-k-\ell-\delta\log_2(q)}$ is small, we have $1-\left(1-2^{v2^a-k-\ell-\delta\log_2(q)}\right)^M\approx M2^{v2^a-k-\ell-\delta\log_2(q)}$, and the cost becomes
$$\frac{\max_{i\in [1 ; a]}\left\{2^{k+\ell+\log_2(q)\big(\delta-\gamma(i)\big)-v\big(2^a-2^i\big)}\right\}}{M}.$$
\end{prop}
For all the cases we consider, the best setting for Wagner's algorithm always results to be that with $a = 1$.
In order to provide a practical evidence of this fact, let us focus (as in \cite{Bricout}) on the case of amortised lists, i.e., the one in which all lists have the same average size, corresponding to that of the initial ones (that is, $2^v$ for some $v\leq \frac{k+\ell}{2^a}$).
It is easy to see that this happens if we choose $u_i = i \frac{v}{\log_2(q)}$, which leads to $\delta = \frac{v\big(2^{a-1}-1\big)}{\log_2(q)}$.
Thus, we estimate the cost of Wagner's algorithm as
\begin{align*}
\frac{2^v}{M2^{v2^a-k-\ell-v\big(2^{a-1}-1\big)}}  = \frac{2^v}{2^{v\big(2^{a-1}+1\big)-k-\ell+\log_2(M)}} = 2^{k+\ell-\log_2(M)-v2^{a-1}}.
\end{align*}
Let us compare this case with that of $a = 1$, with initial lists of sizes $2^{\frac{k+\ell}{2}}$: in such a case, the asymptotic complexity is $2^\frac{k+\ell}{2}$.
Wagner's algorithm with more than one level will then be more convenient if 
$2^{k+\ell-\log_2(M)-v2^{a-1}} < 2^{\frac{k+\ell}{2}}$, 
that is
$$v > \frac{k+\ell}{2^a}-\log_2(M).$$
Remember that it must be $v\leq \frac{k+\ell}{2^a}$: thus, unless $M$ is very large, either there are no values of $v$ for which $a>1$ becomes convenient or, even if they exist, they lead to a really limited advantage in the algorithm cost. 
For instance, if the solution is unique (i.e., if $M = 1$), there are no values of $v$ for which $a>1$ is convenient and the performance of Wagner's algorithm gets worse if $a$ increases over $1$.
\medskip

We remark that the range of convenient values for $v$ is actually thinner, since we are neglecting polynomial factors which increase when $a$ increases.

Based on the above considerations, we are ready to derive a closed formula for the complexity of solving the initial R-SDP instance using Wagner's algorithm on one level.
The whole procedure we consider is reported in Algorithm \ref{wagner_a1}, and its cost 
is detailed in the next proposition.

\begin{algorithm}[ht!]\label{wagner_a1}
\KwIn{$\mathbf{H} \in \mathbb{F}_{q}^{(n-k) \times n}$, $\mathbf{s}\in\mathbb{F}_{q}^{n-k}$, $\ell\in\{1,\ldots,n-k\}$, $v\in \mathbb N$ such that $v\leq \frac{k+\ell}{2}$} 
\KwOut{A $\mathbf{e}\in\{\pm 1\}^n$ such that  $\mathbf{eH}^\top=\mathbf{s}$}
Pick $\pi\rand\mathfrak S_n$. \\Use PGE to transform the initial instance as  $(\6s', \6s'') = \6s\6S$ and $\6S \6H= \begin{bmatrix}\6I_{n-k-\ell} & \6H'\\
\60_{\ell\times(n-k-\ell)} & \6H''\end{bmatrix}$; if it is not possible, restart from line 1. 
\\Choose random subsets $\mathcal{R}_0, \mathcal{R}_1 \subseteq \{\pm 1\}^{(k+\ell)/2}$, each of size $2^v$.
\\Build lists $\mathcal L_0$, $\mathcal L_1$, using the sub-matrix $\6H''$, i.e., 
$$\mathcal L_0 = \left\{(\6z = \6p\6H''^\top_0, \6p)\mid \6p\in\mathcal{R}_0\right\},$$
$$\mathcal L_1 = \left\{(\6z = \6p\6H''^\top_1 -\6s'', \6p)\mid \6p\in\mathcal{R}_1\right\}.$$ \\
\For{$(\6z_1,\6p_1)\in\mathcal L_1$}{
Search for $(\6z_0,\6p_0)\in\mathcal L_0$ such that $\6z_0 +\6z_1 = \60$. Store $(\6p_0, \6p_1) $ in $\mathcal{L}$.} 
\For{$\6e'' \in \mathcal{L}$}{
 compute $\6e' = \6s'-\6e''\6H''^\top$\\
\If{$\6e'\in\{\pm 1\}^{n-k-\ell}$}{ 
\Return $\pi^{-1}\big([\6e', \6e'']\big)$}
}
Restart from line 3. \\
\caption{PGE+SS approach, using Wagner's algorithm with $a = 1$.}
\end{algorithm}

\begin{prop}\label{prop:wagner_1_complexity}
Let us consider an R-SDP instance given by a random $\6H\in\mathbb F_q^{(n-k)\times n}$ with rank $n-k$ and $\6s\rand\left\{\6e\6H^\top \mid \6e\in\{\pm 1\}^n\right\}$. 
We assume that the R-SDP instance given by $(\6H,\6s)$ has $M = 1+2^{n\big(1-(1-R)\log_2(q)\big)}$ solutions.
Then, Algorithm \ref{wagner_a1} finds one of these solutions with the following cost 
\begin{align*} 
C_{\mathrm{PGE}}+\frac{C_{\mathrm{List}}+N_{\mathrm{Test}}C_{\mathrm{Test}}}{1-\big(1-2^{2v-k-\ell}\big)^M},
\end{align*}
where
$$C_{\mathrm{PGE}} = \frac{(n-k-\ell)^2(n-k+1)\left\lceil \log_2(q)\right\rceil^2}{\prod_{j = 1}^{n-k}\left(1-q^{-j}\right)}, \hspace{2mm} C_{\mathrm{Test}} = \frac{q}{q-2} (k + \ell) \lceil \log_2(q) \rceil,$$
$$C_{\mathrm{List}} = 2^{v+1}\left((v+1)+\frac{k+\ell}{2}\ell\left\lceil\log_2(q)\right\rceil\right),$$
$$
N_{\mathrm{Test}} = \big(1-2^{2v-k-\ell}\big)^M2^{2v-\ell\log_2(q)} + \bigg(1-\big(1-2^{2v-k-\ell}\big)^M\bigg)\frac{m'+(2^{2v}-m')q^{-\ell}}{(1+m')},
$$
being
$$\hspace{2mm} m' = M\frac{2^{2v-k-\ell}}{1-\big(1-2^{2v-k-\ell}\big)^M}.$$
\end{prop}
The proof is reported in Appendix B.

\subsection{Considered scenario}\label{sec:scenario}

In the next section we describe our adaptation of the CVE identification scheme to the R-SDP problem, relying on the analysis of the previous sections to devise secure parameters.
In particular, we consider a code with length $n$ and dimension $k$, defined over a finite field with $q\geq 5$ elements, (with $q$ being a prime), described by a random $\6H\in\mathbb F_q^{(n-k)\times n}$ with rank $n-k$. 
In the considered application, the matrix $\6H$ is public, and the target syndrome is obtained as $\6s = \6e\6H^\top$, where $\6e$ is a randomly sampled vector from $\{\pm 1\}^n$.
The vector $\6e$ corresponds to the secret key, while the syndrome $\6s$ is the public key.
Hence, the pair $(\6H,\6s)$ represents an R-SDP instance, and finding a solution to the problem is equivalent to determining either the secret $\6e$ or an equivalent vector. 
The security level of the scheme is given by the cost of solving such an instance of the R-SDP, which allows designing parameters $n$, $k$, and $q$ for a given target security parameter $\lambda$.
\medskip

In particular, we will use Proposition \ref{prop:num_solutions} to estimate the average number of solutions $M$, and consider both the brute force and the PGE+SS approaches to estimate the hardness of recovering the secret $\6e$ from the public key.
Notice that the brute force approach has better performance when the code has small rate (roughly, significantly lower than $1/2$), while for large code rates the PGE+SS approach becomes more convenient. 
We will consider parameters $n$, $k$ and $q$ such that $k/n$ is quite larger than $1/2$ (namely, around $0.8$), and thus we will focus on the PGE+SS approach to estimate the security level, whose complexity can be estimated through Proposition \ref{prop:wagner_1_complexity}.
Obviously, we consider the cost corresponding to the best choice of both parameters $\ell\in\{1,\ldots,n-k\}$ and $v\in\{0,\ldots,\frac{\ell+k}{2}\}$ (i.e., the ones leading to the lowest complexity).

Notice that the security of our scheme is based on the hardness of finding one out of multiple solutions (which, in our case, are given by all vectors that multiplied by $\6H^\top$ result in the public key $\6s$).
A similar setting (which somehow resembles the Decoding One-Out of Many problem \cite{doom}) is employed in other cryptosystems such as WAVE \cite{debris2019wave}. 
However, as an important difference, in our case the expected number of existing solutions $M$ is extremely small.
Indeed, we make use of Proposition \ref{prop:num_solutions} to estimate the number of such equivalent solutions, but choose parameters for which $M$ is only moderately larger than $1$.
In other words, we consider the setting in which the R-SDP admits more than one solution with rather high probability, but limit our interest to the case in which the number of such solutions is quite small. \\

We finally remark that, when the finite field size $q$ is some prime power $q = p^m$, an R-SDP instance may be be mapped into a new instance, defined over some subfield subcode of the code described by $\6 H$.
Indeed, for an integer $ m' \mid m$, we have that $\mathbb F_{p^m}$ can be seen as a vector space over $\mathbb F_{p^{m'}}$ of dimension $m/m'$.
Depending on a choice of basis, one can use an isomorphism projecting each element of $\mathbb F_{p^m}$ into a vector of length $m/m'$ and entries over $\mathbb F_{p^{m'}}$.
By applying this  on $\6 H$ and $\6s$, we obtain a new R-SDP instance with the inputs $\6 H' \in \mathbb F_{p^{m'}}^{(n-k)\frac{m}{m'} \times n}$ and $\6s' \in \mathbb F_{p^{m'}}^{(n-k)\frac{m}{m'}}$. By considering such a subfield subcode, the dimension of the initial code will reduce significantly and thus the new instance has a smaller complexity to be solved.

In order to completely avoid such attacks, in the remaining sections we stick to the case of prime fields $\mathbb{F}_p$. In addition, we avoid finite fields of characteristic 2, for which $-1 =1$ and, with the restriction of a full weight vector $\6e$, it follows that $\6e$ is necessarily the all-one vector.

%% file: wagner.tex
\resizebox{\columnwidth}{!}{
\begin{tikzpicture}

\draw (-0.5,0) rectangle (0.5,1.5);

\node [anchor = west ]at (3,0.75)(L0_label) {Final set of solutions};

\draw[->] (L0_label) edge (0.50,0.75);


\draw (-0.5-2,0-3) rectangle (0.5-2,1.5-3);
\draw[pattern=north east lines, pattern color=gray] (-0.5-2,0-3) rectangle (0.5-2,-3+0.7);
\draw[pattern=north west lines, pattern color=green] (-0.5-2,-3+0.7) rectangle (0.5-2,-3+1.5);

\draw (-0.5+2,0-3) rectangle (0.5+2,1.5-3);
\draw[pattern=north west lines, pattern color=blue] (-0.5+2,0-3) rectangle (0.5+2,-3+0.7);
\draw[pattern=north east lines, pattern color=red] (-0.5+2,-3+0.7) rectangle (0.5+2,-3+1.5);

\draw (-2,-1.5) edge (0,0);
\draw (2,-1.5) edge 
node [anchor = west]{\hspace{5mm}\textcolor{black}{Merge on remaining $\ell-u_0$ positions}}(0,0);


\draw (-0.5-3,0-6) rectangle (0.5-3,1.5-6);
\draw[pattern=north east lines, pattern color=gray] (-0.5-3,0-6) rectangle (0.5-3,-6+0.7);

\draw (-0.5-1,0-6) rectangle (0.5-1,1.5-6);
\draw[pattern=north east lines, pattern color=gray] (-0.5-1,0-6) rectangle (0.5-1,-6+0.7);

\draw (-0.5+1,0-6) rectangle (0.5+1,1.5-6);
\draw [pattern=north west lines, pattern color=blue](-0.5+1,0-6) rectangle (0.5+1,-6+0.7);

\draw (-0.5+3,0-6) rectangle (0.5+3,1.5-6);
\draw [pattern=north west lines, pattern color=blue](-0.5+3,0-6) rectangle (0.5+3,-6+0.7);

\draw node at (-3,-6.5) {$\mathcal L_0^{(0)}$};
\draw node at (-1,-6.5) {$\mathcal L_1^{(0)}$};
\draw node at (1,-6.5) {$\mathcal L_2^{(0)}$};
\draw node at (3,-6.5) {$\mathcal L_3^{(0)}$};

\draw (-3,-4.5) edge 
(-2,-3) ;
\draw (-1,-4.5) edge (-2,-3);
\draw (1,-4.5) edge (2,-3);
\draw (3,-4.5) edge node [anchor = west]{\hspace{5mm}\textcolor{black}{Merge on $u_0$ positions}} (2,-3) ;

\draw[<->] (3.7,-6) to node[anchor = west] {$u_0$} (3.7,-5.3) ;

\draw[<->] (2.7,-3) to node[anchor = west] {$u_0$} (2.7,-2.3) ;
\draw[<->] (2.7,-2.3) to node[anchor = west] {$\ell-u_0$} (2.7,-1.5) ;
\draw[<->] (4.2,-3) to node[anchor = west] {$\ell$} (4.2,-1.5) ;


\draw[thick, gray, dashed] (-6,-1) to (8,-1) ;
\draw[thick, gray, dashed] (-6,-4) to (8,-4) ;

\node at (-6,-4.5){\textbf{Level $0$}};
\node at (-6,-1.5){\textbf{Level $1$}};
\node at (-6,1.5){\textbf{Level $2$}};

\end{tikzpicture}
}

%% file: Previous.tex
In this section we show how new instances of code-based ZK-ID schemes can be built by exploiting the hardness of the R-SDP.
In particular, we focus on the CVE scheme, and revise it using the R-SDP introduced in this paper, as summarized in Fig.~\ref{fig:cayrel_lee}.  For the sake of clarity, we report the procedure for the case of a single round of communication, but we remark that it is always possible to take advantage of the compression technique  described in detail in Fig.~\ref{fig:cayrelcompression}, when multiple rounds are considered.

\begin{figure}[ht!]\small
\centering
\begin{tabular}{p{6cm}p{1cm}p{4.8cm}}
\multicolumn{3}{l}{\textsf{Public Data}\quad Parameters $p,n,k,t\in\9N$, parity-check matrix $\6H\in\mathbb F_p^{(n-k) \times n}$}\\
\multicolumn{3}{l}{\textsf{Private Key}\quad $\6e\in E_{n,p,t}$}\\
\multicolumn{3}{l}{\textsf{Public Key}\quad\ $\6s = \6e\6H^\top\in\mathbb F_p^{n-k}$} \\[5pt]
\hline
\textsf{PROVER} & & \multicolumn{1}{r}{\textsf{VERIFIER}}\\
Choose $\6u\xleftarrow{\$}\mathbb F_p^n$, $\tau\xleftarrow{\$} \widetilde{\mathfrak{M}}_n$ &  & \\
Set $c_0 = \textsf{Hash}\big(\tau,\6u\6H^\top\big)$  &  &\\
Set $c_1 = \textsf{Hash}\big(\tau(\6u), \tau(\6e)\big)$  &  &\\
 & $\xlongrightarrow{(c_0,c_1)}$  &\\
 &&\multicolumn{1}{r}{Choose $z\xleftarrow{\$}\mathbb F_p^*$}\\
&$\xlongleftarrow{z}$&\\
Set $\6y =  \tau(\6u+z\6e)$ &&\\
&$\xlongrightarrow{\6y}$&\\
& & \multicolumn{1}{r}{Choose $b\rand \{0,1\}$}\\
& $\xlongleftarrow{b}$ &\\
If $b=0$, set $f:=\tau$ & & \\
If $b=1$, set $f:=\6e' = \tau(\6e)$ & & \\
&$\xlongrightarrow{f}$&\\
&&\multicolumn{1}{r}{If $b=0$,  accept if}\\
&&\multicolumn{1}{r}{$c_0 = \textsf{Hash}\big(\tau,\tau^{-1}(\6y)\6H^\top -z\mathbf{s}\big)$}\\
&&\multicolumn{1}{r}{If $b=1$, accept if $\widetilde{\mathrm{wt}}(\6e') = t$}\\
&&\multicolumn{1}{r}{and $ c_1 = \textsf{Hash}\big(\6y - z\6e',\6e'\big)$}\\
\hline
\end{tabular}
\caption{Adaptation of the CVE scheme to restricted error vectors.}\label{fig:cayrel_lee}
\end{figure}

In the protocol, we use again $E_{n,p,t}$
to denote the sphere of vectors in $\{0,\pm 1\}^n$ with restricted weight $t$, and we denote by $\widetilde{\mathfrak{M}}_n\subseteq \mathfrak M_n$ the set of monomial transformations whose scaling factors are only $\pm 1$.

Note that the only modifications, with respect to the CVE scheme, are in the fact that we are restricting the secret key and the monomial transformations.
By doing this, we base the security of the protocol on the hardness of the R-SDP, which we have proven to be NP-complete in Section \ref{subsec:NP}.
A formal security analysis of the proposed scheme is provided next.
When designing practical parameters, we suggest to choose $t = n$, since this choice allows to reduce the communication cost.
 
\subsection{Security}

We now show that our protocol satisfies all properties required for a zero-knowledge identification scheme. Note that our proofs are very similar to those in \cite{Cayrel2010}, from which our scheme is adapted.
\medskip

\paragraph*{Completeness} It is easy to show that an honest prover is always successfully verified. In fact, if $b=0$, we have $$\tau^{-1}(\6y)\6H^\top-z \6s=(\6u+z\6e)\6H^\top-z \6s=\6u\6H^\top,$$ and therefore $$\textsf{Hash}\big(\tau,\tau^{-1}(\6y)\6H^\top -z\mathbf{s}\big)$$ matches the commitment $c_0$. Similarly, if $b=1$, we have that $$\widetilde{\mathrm{wt}}(\6e')=\widetilde{\mathrm{wt}}(\tau(\6e))=\widetilde{\mathrm{wt}}(\6e) = t$$ and $$\6y - z\6e'=\tau(\6u)+z\tau(\6e)-z\6e'=\tau(\6u).$$ It follows that $\textsf{Hash}\big(\6y - z\6e',\6e'\big)$ matches the commitment $c_1$, and therefore both conditions are verified.
\medskip

\paragraph*{Zero-Knowledge} To prove this property, we construct a simulator $\mathcal S$, modelled as a probabilistic polynomial-time algorithm, that uses a dishonest verifier $\mathcal A$  
as a subroutine. The goal of such a simulator is to produce a communication record that is indistinguishable from one which would be obtained through an honest execution of the protocol. Note that, since our scheme is a 5-pass protocol, $\mathcal A$ has two strategies for his attack, corresponding to the two interactions with the prover. In the first strategy, which we call $\textsf{ST}_0$, $\mathcal A$ takes as input the prover's commitments $c_0,c_1$ and produces a value $z\in\mathbb F_p^*$. In the second strategy, which we call $\textsf{ST}_1$, $\mathcal A$ takes as input both the commitments $c_0,c_1$ and the first response $\6y$, and generates a challenge $b\in\{0,1\}$.
\\The simulator is constructed as follows. First, pick a random challenge $b\xleftarrow{\$}\{0,1\}$, then:
\begin{itemize}
    \item if $b=0$, choose uniformly at random $\6u$ and $\tau$, then find a vector $\hat{\6e}$ such that $\hat{\6e}\6H^\top=\6s$. No limitation is placed on the restricted value of $\hat{\6e}$; so, this can be accomplished by simple linear algebra. Generate the commitments by setting $c_0 = \textsf{Hash}\big(\tau,\6u\6H^\top\big)$ and picking $c_1$ as a random string of the proper length. Call on $\mathcal A$ with input $c_0,c_1$; $\mathcal A$ will apply $\textsf{ST}_0$ and return a value $z\in\mathbb F_p^*$. Compute $\6y =  \tau(\6u+z\hat{\6e})$ and call on $\mathcal A$ again with input $c_0,c_1,\6y$; this time, $\mathcal A$ will apply $\textsf{ST}_1$ and respond with a bit $\hat b$.
    
     \item if $b=1$, choose again $\6u$ and $\tau$ uniformly at random, then pick a random vector $\hat{\6e}$ of the correct restricted value $t$. Generate the commitments by picking $c_0$ as a random string of the proper length, and setting $c_1 = \textsf{Hash}\big(\tau(\6u), \tau(\hat{\6e})\big)$. As before, call on $\mathcal A$ with input $c_0,c_1$; $\mathcal A$ will apply $\textsf{ST}_0$ and return $z\in\mathbb F_p^*$. Compute again $\6y =  \tau(\6u+z\hat{\6e})$ and call on $\mathcal A$ with input $c_0,c_1,\6y$ to obtain the bit $\hat b$.
     
\end{itemize}
At this point, the simulator has two options. If $\hat b=b$, the simulator halts and produces the communication consisting of $c_0,c_1,z,\6 y,b$ and $f$; otherwise, it restarts the procedure. Note that all the objects comprising the record are distributed uniformly at random. Therefore, on an average of $2N$ rounds, the record produced by $\mathcal S$ is indistinguishable from one which would be produced in an honest execution over $N$ rounds, as we conjectured.
\medskip

\paragraph*{Soundness} We now analyze the cheating probability of an adversary, in this case a dishonest prover $\mathcal A$. We show that such an adversary has a cheating probability that is asymptotically (in $p$) close to $1/2$. 
To this end, we show that $\mathcal A$ can behave in one of two ways, depending on what is the expected challenge value. 
In the first case, which we call $\textsf{ST}_0$, assume without loss of generality that $\mathcal A$ is preparing to receive the challenge $b=0$. Then $\mathcal A$ will choose $\6u$ and $\tau$ uniformly at random, and find a vector $\hat{\6e}$ such that $\hat{\6e}\6H^\top=\6s$, without any limitation on the restricted value. Commitments are generated by setting $c_0 = \textsf{Hash}\big(\tau,\6u\6H^\top\big)$ and picking $c_1$ as a random string of the proper length. Thus, $\mathcal A$ is able to successfully answer the challenge $b=0$, regardless of the value $z$ chosen by the verifier. In fact, the value $\6y =  \tau(\6u+z\hat{\6e})$ and the response $f=\tau$ computed by $\mathcal A$ are enough to pass the verification, since the restricted value of $\hat{\6e}$ is not checked.

In the second case, which we call $\textsf{ST}_1$, $\mathcal{A}$ is instead prepared to receive the challenge $b=1$. In this case, $\mathcal A$ will choose again $\6u$ and $\tau$ uniformly at random, then he will pick a random vector $\hat{\6e}$ of the correct restricted weight $t$. Commitments are generated by picking $c_0$ as a random string of the proper length, and setting $c_1 = \textsf{Hash}\big(\tau(\6u), \tau(\hat{\6e})\big)$. Thus, $\mathcal A$ is able to successfully answer the challenge $b=1$, regardless of the value $z$ chosen by the verifier. In fact, the value $\6y =  \tau(\6u+z\hat{\6e})$ and the response $f=\tau(\hat{\6e})$ computed by $\mathcal A$ are enough to pass the verification, since the same vector $\hat{\6e}$ is used to calculate both objects,  and $\hat{\6e}$ has the correct restricted value.\medskip

Note that the adversary's strategy can be improved in both cases, by taking a guess $\hat z$ on the value $z$ chosen by the verifier, so that $\mathcal A$ is able to answer not only the challenge $b=0$ regardless of $z$, but also the challenge $b=1$ if $z$ was guessed correctly -- or viceversa. With this improvement, we can calculate the probability of success of the adversary as follows, where we model the values $b$ and $z$ as random variables:
\begin{align*} 
 \pr[\7A\text{ is accepted}] &  =\sum_{i=0}^1  \pr[\textsf{ST}=\textsf{ST}_i] (\pr[b=i]    +  \pr[b=1-i] \pr[z=\hat z]) \\
 & =\frac{p}{2(p-1)} .
\end{align*}

To conclude this section, we state the following theorem, relating the cheating probability to the security of the hash function and finding the secret key in the scheme.

\begin{theorem}\label{extractor}
Let $\7V$ be an honest verifier, running $N$ rounds of the protocol in Fig.~\ref{fig:cayrel_lee} with a dishonest prover $\7A$. If $\7A$ is accepted with probability $(\frac{p}{2(p-1)})^N+\epsilon$, where $\epsilon>0$, then it is possible to devise an extractor algorithm $\7E$ that is able to either recover the secret $\6e$, or to find a collision for \textsf{Hash}.
\end{theorem}

The proof of Theorem~\ref{extractor} proceeds along the lines of that given in~\cite[Theorem 2]{Cayrel2010}, and therefore we do not repeat it here for the sake of brevity.

\subsection{Communication cost}

In our scheme, the public matrix is the parity-check matrix of a linear code over $\mathbb F_p$, with length $n$ and dimension $k$.
To reduce the computational complexity of the protocol, we can rely on the systematic form of such a matrix, i.e., we can choose $\6H = [\6I_{n-k} \ \6P]$, where $\6P\in\mathbb F_p^{(n-k) \times k}$.
Note that, since the code is chosen uniformly at random, its full representation is provided by the associated seed.

The public key is the syndrome of the secret key through $\6H$, thus it is a vector of length $n-k$ over $\mathbb F_p$; thus, its representation requires $(n-k)\left\lceil\log_2(p)\right\rceil$ bits.
\medskip

To properly calculate the communication cost of the scheme, we make the following considerations: 
\begin{enumerate}
    \item[-] The vector $\6y$ is random over $\mathbb F_p^n$ and is represented through $n\left\lceil\log_2(p)\right\rceil$ bits.
    \item[-] When $b=0$, the monomial transformation can be represented through the associated seed.
    \item[-] When $b=1$, $\6e'$ is a random vector over $\{0,\pm 1\}^n$ with restricted weight, or equivalently Hamming weight, $t$. 
    In particular, we will consider the worst case of $t = n$, in which $\6e'$ can be efficiently represented by a binary string of length $n$.
\end{enumerate}\medskip

Given these considerations, and assuming $N$ rounds are performed with the compression technique illustrated in Fig. \ref{fig:cayrelcompression}, the average communication cost of our scheme is derived as
$$l_{\textsf{Hash}} + N  \bigg(\left\lceil\log_2(p-1)\right\rceil+n\left\lceil\log_2(p)\right\rceil+1+l_{\sf{Hash}}+\frac{n+l_{\sf{Seed}}}{2}\bigg).$$
For the maximal communication cost, we instead have
$$l_{\textsf{Hash}} + N  \bigg(\left\lceil\log_2(p-1)\right\rceil+n\left\lceil\log_2(p)\right\rceil+1+l_{\sf{Hash}}+\max\{n\hspace{1mm},\hspace{1mm}l_{\sf{Seed}}\}\bigg).$$
To reach a cheating probability not larger than $2^{-\tau}$, the number of rounds is obtained as
$$N = \left\lceil \frac{-\tau}{\log_2\left(\frac{p}{2(p-1)}\right)}\right\rceil.$$
Note that, in practice, this means that the number of rounds is approximately equal to $\tau$. 
\subsection{Practical instances}
\label{subsec: PRIN}

In this section we propose some practical instances of our scheme, and compare them with other code-based identification schemes based on the Hamming metric, at the same security level.
We first briefly describe how secure parameters for the scheme can be designed, by recalling the analysis in Section \ref{sec:scenario}.
We consider parameters $p,n,k$ for the public code (i.e., the code defined by the public parity-check matrix $\6H$) such that solving the R-SDP for an error vector of weight $n$ requires at least $2^\lambda$ operations, where $\lambda$ is the security level expressed in bits. 
As in Proposition \ref{prop:num_solutions}, we estimate the number of solutions of the R-SDP as $M = 1 + 2^{n\big(1-(1-k/n)\log_2(p)\big)}$, and rely on Proposition \ref{prop:wagner_1_complexity} to estimate the complexity of attacks based on the PGE+SS approach.

In particular, we focus on those parameters for which the value of $M$ is particularly small and, de facto, only slightly larger than $1$.

In order to provide a first and direct comparison with existing code-based identification schemes, we consider the same setting as in \cite{Cayrel2010}.
Hence, we fix a security level equal to $\lambda = 87$ bits and a target cheating probability equal to $2^{-16}$, using  hashes and seeds of length $128$ and $160$ bits, respectively.
For the classical CVE scheme, we consider the parameters provided in \cite{Cayrel2010}.
For the sake of comparison, in the appendix we also design updated parameters for the AGS scheme, to target the same security level (the cheating probability of a single round has been conservatively approximated to $1/2$).
For our variant of the CVE scheme, we have $p = 29$, $n = 167$, $k = 132$, for which
\begin{enumerate}
    \item[-] the expected number of solutions is $M = 1.527$;
    \item[-] the PGE+SS attack is optimized by choosing $\ell = 15$, $v = 73$, with a resulting cost of $2^{87.126}$;
    \item[-] the required number of rounds is $N = 17$.
\end{enumerate}\medskip

Table \ref{tab:comparison_87} compares the performance of these three schemes, also taking into account the compression technique. 
\begin{table}[h!]
    \centering
     \caption{Comparison between ZK-ID schemes, for a security parameter $\lambda = 87$ and a cheating probability $2^{-16}$,  assuming seeds and hashes of, respectively, $128$ and $160$ bits.} 
    \begin{tabular}{|c|c|c|c|}\hline
    & CVE & AGS & Rest. CVE \\\hline
    Number of rounds & 17 & 16 & 17 \\
    Public key size (bits) & 512 & 1094 & 175 \\
    Total average comm. cost (kB) & 3.472 & 3.463 & 2.389  \\
    Total max comm. cost (kB) & 4.117 & 4.894 & 2.430 \\\hline
    \end{tabular}
    \label{tab:comparison_87}
\end{table}
As we see, our scheme (denoted as Rest. CVE in the table) yields significant improvements in the communication cost.
As another important advantage, the size of the public key is strongly reduced as well.\medskip

\subsubsection*{Signatures}

Signature schemes can be obtained, in the Random Oracle Model, by applying the well-known Fiat-Shamir transform~\cite{fiat1986prove} to any ZK-ID. The transform is very intuitive for 3-pass schemes, in which the protocol is made non-interactive by generating the challenge bits as the hash output of the commitment and the message. The idea can easily be generalized to 5-pass schemes such as ours, as illustrated in~\cite{cayrelextended}; in this case, the communication cost roughly corresponds to the size of a signature. Some minor optimizations are possible, but, especially for the case of 5-pass schemes, lead only to a very limited improvement. Thus, to keep the analysis as simple as possible, we do not consider such optimizations. Note that, for a signature scheme to be of practical interest, the requirements are higher in terms of security with respect to those considered in the previous comparison.
Thus, we provide below parameters for $\lambda = 128$, corresponding to an authentication level of $2^{-128}$, and we update the lengths of both seeds and hash digests, fixing $l_{\sf{Seed}} = l_{\sf{Hash}} = 256$. 

For our scheme, we recommend to choose $p = 31$, $n = 256$, $k = 204$, for which
\begin{enumerate}
    \item[-] the expected number of solutions is $M = 1.326$;
    \item[-] the PGE+SS attack is optimized by choosing $\ell = 22$, $v = 13$, with a resulting cost of $2^{128.029}$;
    \item[-] the required number of rounds is $N = 135$.
\end{enumerate}
\medskip
In Table \ref{tab:comparison_128} the features of our scheme are compared, again, with those of CVE and AGS.

\begin{table}[h!]
    \centering
    \caption{Comparison between signature schemes derived from ZK-ID schemes with $128$-bit security, considering seeds and hashes of $256$ bits.}
    \begin{tabular}{|c|c|c|c|}\hline
    & CVE & AGS & Rest. CVE \\\hline
    Number of rounds & 129 & 128 & 135\\
    Public key size (bits) & 832 & 1574 & 260\\
    Average sig. size (kB) & 43.263 & 41.040 & 30.373 \\
    Max sig. size (kB) & 51.261 & 56.992 & 30.373\\\hline
    \end{tabular}
    \label{tab:comparison_128}
\end{table}

We observe that, also in this case, our scheme achieves significant reductions in all the considered sizes over alternative solutions.

To complete the picture, we comment about some schemes that appeared recently in literature. First, we consider the work of \cite{Bellini2019}, that is an adaptation of Veron's scheme \cite{Veron97} to the rank metric. For the instance denoted as cRVDC-125 in the paper, which reaches a security of $125$ bits, the average signature size is estimated as $22.482$ kB and the public key is $1212$ bits long. Note that, despite a slightly larger security level, our scheme leads to larger signature sizes than those of cRVDC-125, but it exhibits much more compact public keys. Next, we consider Durandal~\cite{Durandal}, which is again obtained via Fiat-Shamir and also uses the rank metric, but is based on a different ZK-ID that is an adaptation of the Schnorr-Lyubashevsky approach~\cite{lyubashevsky2012lattice}. The authors propose two sets of parameters: for the smallest of the two, the public key size is $121 961$ bits and the signature size is $32 514$ bits, corresponding to approximately $15$ kB and $4$ kB, respectively. It is immediate to notice that the main benefit of this approach is the very short signature size, due to the absence of soundness error, meaning that no repetitions of the protocol are necessary. However, this comes at the cost of a considerably larger public key (as well as an ad-hoc security reduction and other similar security concerns). LESS~\cite{less} is an innovative scheme based on an alternative approach, which exploits the code equivalence problem rather than the hardness of decoding. The scheme, after revising its parameters due to a drastic improvement of the known solvers~\cite{ward}, presents similar sizes for both public keys and signatures, around the $15$ kB mark. Thus, similar to Durandal, the signature compares favorably to ours, but the public key is of a much bigger scale. Finally, Wave~\cite{debris2019wave} uses a completely different paradigm (hash-and-sign), which is not based on ZK-ID. It follows that the differences in performance with our scheme are even starker. In fact, the protocol uses random linear codes in the Hamming metric, leading to a public key of  $\mathcal{O}(n^2)$ bits and a signature of $\mathcal{O}(n)$ bits, where $n$ is the length of the chosen linear code. The authors set this parameter at $n=8492$, which leads to very unbalanced sizes, roughly $3.2$ MB for the public key, and about $1.6$ kB for the signature.

\subsection{Implementation aspects}

Given that the prover and the verifier perform, besides hash function computation, only basic linear algebra operations (i.e., sums, multiplications and monomial transformations), we expect our protocol to be at least as fast as the other ZK-ID schemes we have considered. 
In particular, with respect to the standard CVE scheme, it is very likely that our solution can bring important benefits on the implementation side.
In fact, our scheme uses codes with essentially the same length and dimension, but in a finite field of smaller size: given that sums and multiplications in $\mathbb F_p$ essentially cost $O\big(\left\lceil \log_2(p)\right\rceil \big)$ and $O\big(\left\lceil \log_2(p)\right\rceil^2  \big)$, respectively, a smaller finite field leads to a simplified and faster algebra.

Furthermore, restricted monomial transformations are easier to handle, with respect to the general case of monomial transformations over $\mathbb F_p$.
In fact, multiplying by $\pm 1$ either corresponds to doing nothing, or simply performing a sign change. Roughly speaking, scaling according to a restricted transformation may cost as much as $n$ sums, instead of $n$ multiplications.
Finally, note that computing the inverse of a restricted monomial transformation is also easier: indeed, the inverse of $\pm 1$ is equal to itself (so, no actual inverse needs to be computed).
Given all these considerations, we believe that an optimized, ad-hoc implementation of our scheme can achieve particularly favourable running times.
\medskip

To provide some preliminary measure, we have implemented our scheme using Sagemath; the corresponding code is open source and available online\footnote{The proof-of-concept implementation of our scheme is available at \url{https://re-zkid.github.io/}.}. 
We have performed experiments on an Intel(R) Core(TM) i7-8565U CPY, running at $1.80$ GHz, for the instances reaching $128$ bits of security. 

We have averaged over $1000$ runs, obtaining a time of $4.78$ ms for a single round verification. 
For  the  sake  of  completeness, an implementation for the case of multiple  rounds is also publicly available, in which we have used the compression technique to reduce the communication cost. 
We remark that these implementations may be strongly optimized and there is still large room for improvements.

%% file: Concl.tex
In this paper we have studied generalizations of some decoding problems, and their application to zero-knowledge code-based identification schemes.
In particular, we have  introduced the R-SDP, a new decoding problem in which the searched error, corresponding to the given syndrome, must have entries belonging to a restricted subset of the finite field.
We have shown that the decisional version of this new problem is NP-complete, via a reduction from the Hamming version of the SDP, and have adapted classical arguments about random codes (such as the Gilbert-Varshamov bound) to take into account error vectors with this particular structure.
We have assessed the complexity of solving the R-SDP adapting modern techniques. 
We have provided an adaption of the CVE scheme to the case of restricted error vectors and compared this proposal to the original CVE scheme and to the AGS scheme. Finally, we have observed that using restricted error vectors we can achieve a reduction in the communication cost of more than $25\%$ over classical approaches, which coincides with the achievable reduction in the signature size when these schemes are used as the basis for digital signature schemes obtained through the Fiat-Shamir transform.

%% file: AppendixA.tex
The AGS scheme \cite{Aguilar2011} is constructed upon quasi-cyclic codes over $\mathbb F_2$. 
Let us consider a vector $\6a\in\mathbb F_2^{jk}$ divided into $j$ blocks of $k$ entries each, that is,
\[
\6a=[a^{(0)}_0,\ldots,a^{(0)}_{k-1} | \ldots | a^{(j-1)}_0,\ldots,a^{(j-1)}_{k-1}].
\]
We use $\rho_i^{(k)}$ to denote a function that performs a block-wise cyclic shift of $\6a$ by $i$ positions towards right, i.e.,
$$\rho_i^{(k)}(\6a)=\left[ a^{(0)}_{-i \mod k},\ldots, a^{(0)}_{k-1-i \mod k}| \ldots| a^{(j-1)}_{-i \mod k},\ldots, a^{(j-1)}_{k-1-i \mod k} \right].$$

\begin{figure}[ht!]
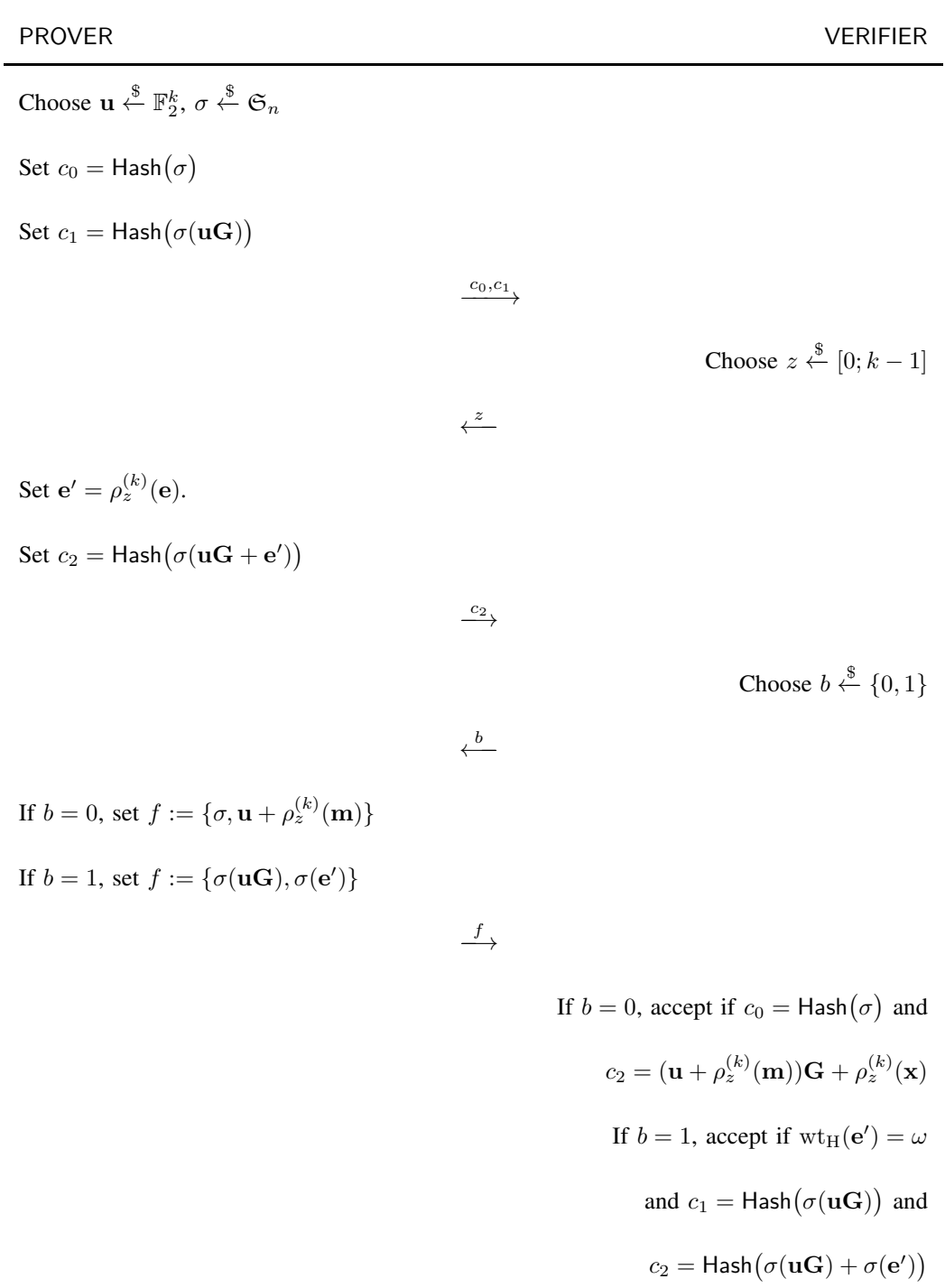
\small\centering
\begin{tabular}{p{6cm}p{1cm}p{4.8cm}}
\multicolumn{3}{l}{\textsf{Public Data}\quad Parameters $k,n,\omega\in\9N$, hash function $\textsf{Hash}$, generator matrix $\6G\in\mathbb F_2^{k\times n}$}\\
\multicolumn{3}{l}{\textsf{Private Key}\quad $\6m\in\mathbb F_2^k$, $\6e\in \mathrm{S}^{\mathrm H}_{n,2,\omega}$}\\
\multicolumn{3}{l}{\textsf{Public Key}\quad $\6x=\6m\6G+\6e\in \mathbb F_2^n$} \\[5pt]
\hline
\textsf{PROVER} & & \multicolumn{1}{r}{\textsf{VERIFIER}}\\
\hline
Choose $\6u\xleftarrow{\$}\mathbb F_2^k$, $\sigma\xleftarrow{\$} \mathfrak S_n$  &  & \\
Set $c_0 = \textsf{Hash}\big(\sigma\big)$  &  &\\
Set $c_1 = \textsf{Hash}\big(\sigma(\6u\6G)\big)$  &  &\\
&$\xlongrightarrow{c_0,c_1}$&\\
& & \multicolumn{1}{r}{Choose $z\xleftarrow{\$}[0;k-1]$}\\
&$\xlongleftarrow{z}$&\\
Set $\6e'=\rho_{z}^{(k)}(\6e)$.\\
Set $c_2=\textsf{Hash}\big(\sigma(\6u\6G+\6e')\big)$\\
&$\xlongrightarrow{c_2}$&\\
& & \multicolumn{1}{r}{Choose $b\rand \{0,1\}$}\\
& $\xlongleftarrow{b}$ &\\
If $b=0$, set $f:=\{\sigma,\6u+\rho^{(k)}_{z}(\6m)\}$ & & \\
If $b=1$, set $f:=\{\sigma(\6u\6G),\sigma(\6e')\}$ & & \\
&$\xlongrightarrow{f}$&\\
&&\multicolumn{1}{r}{If $b=0$, accept if $c_0 = \textsf{Hash}\big(\sigma\big)$ and}\\
&&\multicolumn{1}{r}{$c_2= (\6u+\rho^{(k)}_{z}(\6m))\6G+\rho^{(k)}_{z}(\6x)  $}\\
\multicolumn{3}{r}{If $b=1$, accept if $\mathrm{wt}_{\mathrm H}(\6e') = \omega$}\\
&&\multicolumn{1}{r}{and $ c_1=\textsf{Hash}\big(\sigma(\6u\6G)\big)$ and  }\\
&&\multicolumn{1}{r}{$c_2= \textsf{Hash}\big(\sigma(\6u\6G)+\sigma(\6e')\big)$}\\
\hline
\end{tabular}
\caption{The AGS scheme.}
\label{fig:aguilarcompre}
\end{figure}

The AGS scheme is described in Fig. \ref{fig:aguilarcompre}. In such a scheme, the cheating probability asymptotically tends to $\frac{1}{2}$ \cite{Aguilar2011}. 
However, in \cite{Aguilar2011} a direct expression for the actual cheating probability is not provided, thus we conservatively assume that its value is  $\frac{1}{2}$, which is optimal.   
When performing $N$ rounds, the average communication cost is
$$l_{\textsf{Hash}} + N\bigg(\left\lceil\log_2(k)\right\rceil+1+2l_{\textsf{Hash}}+\frac{l_{\textsf{Seed}}+k+n+\psi(n,\omega,2)}{2}\bigg),$$
while the maximum communication cost is
$$l_{\textsf{Hash}} + N\bigg(\left\lceil\log_2(k)\right\rceil+1+2l_{\textsf{Hash}}+\min\{l_{\textsf{Seed}}+k\hspace{1mm},\hspace{1mm}n+\psi(n,\omega,2)\}\bigg).$$
In \cite{Aguilar2011}, three parameters sets are proposed:
\begin{enumerate}
    \item[-] $n=698$, $k=349$, $\omega=70$, for $81$-bits security;
    \item[-] $n=1094$, $k=547$, $\omega=109$, for $128$-bits security. 
\end{enumerate}
Taking into account advances in binary ISD techniques \cite{bjmm}, as well as the polynomial gain due to the quasi-cyclic structure of the codes \cite{doom}, the security level of this scheme can be approximately estimated as $\omega-\frac{1}{2}\log_2(k)$ bits.
Such a security level is below the one originally estimated in \cite{Aguilar2011}.
Thus, we have updated the scheme parameters as follows, in order to reach security levels that can directly be compared with those achieved by the CVE scheme in \cite{Cayrel2010}:
\begin{enumerate}
    \item[-] $n = 1094$, $k=547$, $\omega = 92$, for $87$-bits security;  
    \item[-] $n = 1574$, $k=787$, $\omega=133$, for $128$-bits security. 
\end{enumerate}

%% file: AppendixB.tex
In this Appendix we derive the cost of Algorithm \ref{wagner_a1}, used to solve a random R-SDP instance represented by $\6H\in\mathbb F_q^{(n-k)\times n}$ with rank $n-k$ and $\6s\rand\left\{\6e\6H^\top\mid\6e\in\{\pm 1\}^n\right\}$.

\begin{enumerate}
\item 
First, the probability that the PGE step in line 2 is successful is given by $ \prod_{j = 1}^{n-k} \left(1-q^{-j}\right)$, which  is the probability for a matrix of size $(n-k)\times (n-k)$ to have full rank $n-k$. 

We estimate the cost of performing the PGE step as $(n-k-\ell)^2(n-k+1)\left\lceil \log_2(q)\right\rceil^2$, hence, to execute the instructions in lines 1 and 2 in the algorithm, we have an average cost of 
$$C_{\mathrm{PGE}} = \frac{(n-k-\ell)^2(n-k+1)\left\lceil \log_2(q)\right\rceil^2}{\prod_{j = 1}^{n-k} \left(1-q^{-j}\right)}.$$

\item
Remember that $\6H''$ has dimensions $\ell\times (k+\ell)$, and that we split it into two sub matrices $\6H''_0$ and $\6H''_1$, each with dimensions $\ell\times \frac{k+\ell}{2}$.
To build $\mathcal L_0$ and $ \mathcal{L}_1$, we generate $L = 2^v$ random vectors $\6p$ of length $\frac{k+\ell}{2}$ and compute (and store) the corresponding $\6z = \6p\6H''^\top_0$, respectively $\6z = \6p \6H''^\top_1 - \6s''$. 
We assume that the cost to create each list element is identical to that of computing $\6z$, which requires the addition of $(k+\ell)/2$ vectors with length $\ell$ and entries over $\mathbb F_q$, and hence is given by $\frac{k+\ell}{2}  \ell\left\lceil\log_2(q)\right\rceil$ binary operations. 
Since $|\mathcal L_i| = 2^v$, the overall cost of producing $\mathcal L_i$ can be estimated as
$2^v\frac{k+\ell}{2}\ell\left\lceil\log_2(q)\right\rceil.$ 
In line 5 to 6 we search for all $(\6z_0,\6p_0) \in \mathcal{L}_0$ and $(\6z_1, \6p_1) \in \mathcal{L}_1$, such that $\6z_0+\6z_1=\60$. We can do this by joining and sorting the two lists, with a cost of $2L \log_2(2L) = (v+1)2^{v+1}$ operations. The cost of finding pairs of vectors from both lists that sum to zero, i.e., elements in $\mathcal{L}$, is negligible, since it can be done during sorting.
Hence, both lists are generated, sorted and merged with a cost of
$$C_{\mathrm{List}} = (v+1)2^{v+1} + 2^{v+1}\frac{k+\ell}{2}\ell\left\lceil\log_2(q)\right\rceil = 2^{v+1}\left(v+1+\frac{k+\ell}{2}\ell\left\lceil\log_2(q)\right\rceil\right).$$ 
Note that the list $\mathcal L_1$ can also be generated on the run, which leads to a polynomial decrease in the complexity analysis by a factor between $1$ and $2$. 

\item
We rely on Proposition \ref{prop:pr_wagner} to estimate the probability that the final list $\mathcal L$ contains at least a vector leading to a solution for the initial R-SDP instance.
In this case we have $\delta = 0$ (since we do not filter any of the candidates in $\mathcal R_0$, $\mathcal R_1$), hence we have a success probability of
$$1-\big(1-2^{2v-k-\ell}\big)^M.$$
In particular, the reciprocal of 
the above corresponds to the average number of times we repeat the instructions in lines 3--10. 
\item 
We now call an element of $\mathcal L$ a \emph{valid solution} if it leads to a solution of the initial R-SDP. 
We observe the following: If $(\6z_0,\6p_0) \in \mathcal{L}_0$ and $(\6z_1, \6p_1) \in \mathcal{L}_1$ lead to a valid solution, then $(\6p_0,\6p_1)$ will be in the merged list $\mathcal L$. 
Therefore, the probability that $\mathcal L$ contains $m$ valid solutions is equal to the probability that $\mathcal L_0 \times \mathcal L_1$ contains $m$ valid solutions, which can be estimated as
$$\binom{M}{m}\left(2^{2v-k-\ell}\right)^m\left(1-2^{2v-k-\ell}\right)^{M-m}.$$

When there are $m$ valid solutions, the number of elements of $\mathcal L$ that do not lead to a solution for the initial R-SDP instance, can be derived as
$$\frac{2^{2v}-m}{q^\ell},$$ 
hence the average size of $\mathcal L$ is given by $m+\frac{2^{2v}-m}{q^\ell}$.

\item
At this point, one of the following two conditions may occur:
\begin{enumerate}
    \item if $\mathcal L$ does not contain any valid solution, one will test all the candidates in $\mathcal L$ before restarting from line 3.
    This happens with probability $\Pr[m = 0] = \big(1-2^{2v-k-\ell}\big)^M$, and the size of $\mathcal L$ can be estimated as $2^{2v - \ell\log_2(q)}$. 
    \item if $\mathcal L$ contains $m\geq 1$ valid vectors, when scanning the elements in $\mathcal L$, we will at some point find a solution to the initial R-SDP instance (i.e., a vector $\6e'$ satisfying the requirement in Line 9 of the Algorithm). 
    We can derive the expected number $m'$ of valid solutions, conditioned to the fact that $m\geq 1$, as 
    \begin{align*}
    m' = \frac{M2^{2v-k-\ell}}{\Pr[m\geq 1]} = M\frac{2^{2v-k-\ell}}{1-\big(1-2^{2v-k-\ell}\big)^M}.    
    \end{align*}
    Notice that, in this case, $\mathcal L$ contains on average $m'+(2^{2v}-m')q^{-\ell}$ elements. 
    Hence, the number of tests before we find a valid solution can be estimated as 
    $$\frac{m'+(2^{2v}-m')q^{-\ell}}{(1+m')}.$$
\end{enumerate}
To resume, the average number of performed tests before one solution is found can be obtained as
\begin{align*}
N_{\mathrm{Test}} & \nonumber = \Pr[m = 0]2^{2v-\ell\log_2(q)} + \Pr[m\geq1]\frac{m'+(2^{2v}-m')q^{-\ell}}{(1+m')}.
\end{align*}
For each test, we consider a cost given by $C_{\mathrm{Test}} = \frac{q}{q-2} (k + \ell) \lceil \log_2(q) \rceil$, due to early abort.
Since the success probability is given by $\Pr[m\geq1]$, we finally derive the cost of instructions from line 3 to line 11 as
\begin{align*}
\frac{C_{\mathrm{List}}+N_{\mathrm{Test}}C_{\mathrm{Test}}}{\Pr[m\geq1]}.    
\end{align*}
 \end{enumerate}